\documentclass[11pt]{article}
\usepackage[dvipsnames]{xcolor}
\RequirePackage{doi}
\usepackage{hyperref}

\hypersetup{
   colorlinks,
   linktocpage,
    linkcolor=Maroon,
    filecolor=Maroon,      
    urlcolor=Maroon,
    citecolor=Maroon
}
\usepackage{slashed}
\usepackage[utf8]{inputenc}
\usepackage{amsmath, amssymb, amsthm, tabto, epigraph, mathtools, dsfont, verbatim, slashed, mathrsfs}
\usepackage{graphicx, wrapfig}
\usepackage{enumitem}
\usepackage{relsize}
\usepackage{subcaption}
\usepackage{csquotes}
\usepackage{bbm}
\usepackage{tabularx}
\usepackage{comment}

\usepackage[left=2.5cm,right=2.5cm,top=2.5cm,bottom=3cm]{geometry}
\usepackage{tikz}
\usetikzlibrary{arrows,decorations.markings,cd}
\tikzset{
->-/.style={decoration={markings, mark=at position .5 with {\arrow[scale=1.5]{stealth}}}, postaction={decorate}}
}
\usepackage{epsfig}
\usepackage{amssymb}
\usepackage{amsfonts}
\usepackage{amsbsy}
\usepackage{amsmath}
\usepackage{dsfont}
\usepackage{bbm}
\usepackage{upgreek}
\usepackage{amscd}
\usepackage[nosort]{cite}
\usepackage{graphicx}
\usepackage{mathrsfs}
\usepackage{amsmath,amsthm}
\usepackage{slashed}	
\usepackage{dsfont}
\usepackage[utf8]{inputenc}
\numberwithin{equation}{section}

\makeatletter
\DeclareFontFamily{OMX}{MnSymbolE}{}
\DeclareSymbolFont{MnLargeSymbols}{OMX}{MnSymbolE}{m}{n}
\SetSymbolFont{MnLargeSymbols}{bold}{OMX}{MnSymbolE}{b}{n}
\DeclareFontShape{OMX}{MnSymbolE}{m}{n}{
    <-6>  MnSymbolE5
   <6-7>  MnSymbolE6
   <7-8>  MnSymbolE7
   <8-9>  MnSymbolE8
   <9-10> MnSymbolE9
  <10-12> MnSymbolE10
  <12->   MnSymbolE12
}{}
\DeclareFontShape{OMX}{MnSymbolE}{b}{n}{
    <-6>  MnSymbolE-Bold5
   <6-7>  MnSymbolE-Bold6
   <7-8>  MnSymbolE-Bold7
   <8-9>  MnSymbolE-Bold8
   <9-10> MnSymbolE-Bold9
  <10-12> MnSymbolE-Bold10
  <12->   MnSymbolE-Bold12
}{}

\let\llangle\@undefined
\let\rrangle\@undefined
\DeclareMathDelimiter{\llangle}{\mathopen}%
                     {MnLargeSymbols}{'164}{MnLargeSymbols}{'164}
\DeclareMathDelimiter{\rrangle}{\mathclose}%
                     {MnLargeSymbols}{'171}{MnLargeSymbols}{'171}
\makeatother

\DeclareMathOperator{\id}{\mathds{1}}

\DeclareFontShape{OT1}{cmr}{mx}{n}%
    {<->cmr10}{}
\newcommand{\mytitlefont}{\fontseries{mx}\selectfont}
\DeclareMathAlphabet{\titlemath}{OT1}{cmr}{mx}{n}

\def\ie{\begin{equation}\begin{aligned}}
\def\fe{\end{aligned}\end{equation}}

\begin{document}

\begin{titlepage}

\begin{center}

~\\[2cm]

{\fontsize{20pt}{0pt} \mytitlefont  Symmetry Enriched $c$-Theorems \& SPT Transitions}

~\\[0.1cm]

 Clay C\'{o}rdova\footnote{\href{clayc@uchicago.edu}{clayc@uchicago.edu}} and Diego Garc\'{i}a-Sep\'{u}lveda\footnote{\href{dgarciasepulveda@uchicago.edu}{dgarciasepulveda@uchicago.edu}} 

~\\[0.1cm]

{\it Kadanoff Center for Theoretical Physics \& Enrico Fermi Institute, University of Chicago}\\[4pt]

~\\[15pt]

\end{center}

\noindent 
We derive universal constraints on $(1+1)d$ rational conformal field theories (CFTs) that can arise as transitions between topological theories protected by a global symmetry.  The deformation away from criticality to the trivially gapped phase is driven by a symmetry preserving relevant deformation and under renormalization group flow defines a conformal boundary condition of the CFT.  When a CFT can make a transition between distinct trivially gapped phases the spectrum of the CFT quantized on an interval with the associated boundary conditions has degeneracies at each energy level. Using techniques from boundary CFT and modular invariance, we derive universal inequalities on all such degeneracies, including those of the ground state.  This establishes a symmetry enriched $c$-theorem, effectively a lower bound on the central charge which is strictly positive, for this class of CFTs and symmetry protected flows.  We illustrate our results for the case of flows protected by $SU(M)/\mathbb{Z}_{M}$ symmetry. In this case, all SPT transitions can arise from the WZW model $SU(M)_{1},$  and we develop a dictionary between conformal boundary conditions and relevant operators.   

\vfill

\begin{flushleft}
October 2022
\end{flushleft}

\end{titlepage}

\setcounter{tocdepth}{3}

\tableofcontents

\newpage

\section{Introduction}

In this paper we explore the properties of two-dimensional conformal field theories that appear as second order transitions between trivially gapped phases.  For rational CFTs, we derive a universal inequality relating the associated ground state degeneracy of the CFT defined on a strip and the bulk central charge.  We apply this bound to a variety of renormalization group flows enriched by global symmetries where the non-trivial ground state degeneracy is enforced by a symmetry protected trivially gapped phase.

\subsection{Symmetry Enriched Renormalization Group Flows}\label{sec:symflow}

One of the foundational results about quantum field theory is the monotonicity of the renormalization group flow.  In two spacetime dimensions this idea is famously quantified by the $c$-theorem \cite{Zamolodchikov:1986gt}: along a renormalization group flow the central charge $c$ decreases.  In particular, for a conformal field theory which can flow to a gapped phase (which has vanishing central charge) we deduce that the central charge is non-negative
\begin{equation}\label{cpos}
c\geq 0~.    
\end{equation}
This simple result also follows directly from unitarity and reflects a basic feature of Virasoro representation theory.  By now, these monotonicity results have been generalized to different spacetime dimensions \cite{Jafferis:2011zi, Intriligator:2003jj, Casini:2004bw,Casini:2006es,Casini:2015woa,Casini:2017vbe,Komargodski:2011vj,Komargodski:2011xv,Cordova:2015vwa,Cordova:2015fha} and may be derived from a variety of disparate theoretical techniques including anomalies, information theory, and holography---each leading to the same underlying conclusions about coarse graining in field theory.

In the spirit of the ideas to follow, one may also view the inequality \eqref{cpos} as a basic statement about second order phase transitions.  Indeed, intuitively a gapped phase is the long-distance limit of a generic quantum field theory, and may be thought of generally as a topological field theory with only long-range correlation functions but no non-trivial local degrees of freedom.  Meanwhile, a conformal field theory, a gapless system, results from tuning parameters to close the gap and requires non-trivial power law correlation functions measured by $c$. 

In this paper, we will enrich this paradigm by considering flows that preserve an (ordinary) global symmetry $G$.  One of our main aims is to strengthen the bound \eqref{cpos} by tracking the symmetry action to long distances.  We therefore consider a flow triggered by a $G$ invariant operator $\mathcal{O}$
\begin{equation}\label{relopdef}
    \delta S= \lambda \int d^{2}x \ \mathcal{O}(x)~.
\end{equation}
In the simplest analysis, we assume that the infrared is trivially gapped, so in particular $G$ is not spontaneously broken.  In this situation the IR is described by a symmetry protected topological phase (SPT).  In field theoretic language such a model describes a theory of local contact terms for the operators defining the $G$ symmetry.  If we introduce background gauge fields $A$ sourcing $G$, the IR partition function is then a phase, which is specified by a local action
\begin{equation}\label{SPTacti}
    Z_{IR}[A]=\exp\left(2\pi i \int_{X}\omega(A)\right)~,
\end{equation}
where above, $\omega(A)$ may be viewed as the Lagrangian for the SPT and $X$ is the two-dimensional spacetime manifold.  The action \eqref{SPTacti} is non-trivial because it may not be continuously deformed to the trivial action while preserving the energy gap above the unique ground state. 

Symmetry preserving renormalization group flows to trivially gapped phases are common in the study of field theory; however, considered in isolation they cannot strengthen the basic inequality \eqref{cpos}.  The reason is instructive: in defining the UV field theory, we are generally agnostic to the choice of scheme.  The difference between any two schemes is given by a local classical action for the sources, of which \eqref{SPTacti} is an example. Therefore, by adjusting the ultraviolet scheme we are free to assume that the IR partition function resulting from any fixed flow to a trivially gapped phase is described by a completely trivial $G$-SPT.
  
\begin{figure}[t]
        \centering
        \includegraphics[width=0.8 \textwidth]{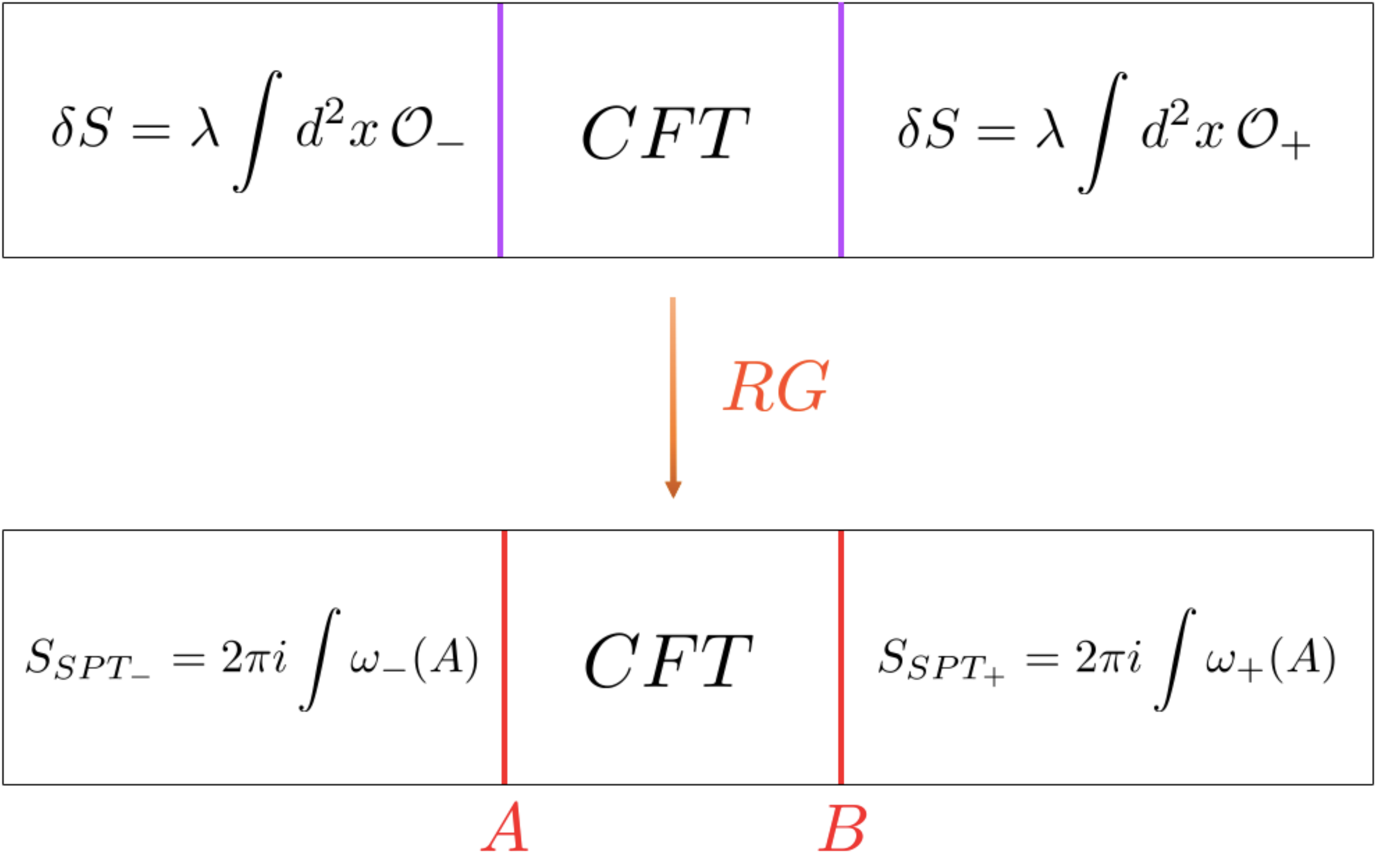} % second figure itself
        \caption{A CFT is deformed in the region $x\ll 0$ by a (symmetry-preserving) relevant operator $\mathcal{O}_{-}$ and in the region $x\gg 0$ by the relevant operator $\mathcal{O}_{+}$. If the IR is trivially gapped, the relevant deformations flow to conformal boundary conditions $A$ and $B$, and we obtain the CFT on a strip.  The trivially gapped regions with $|x|\gg0$ are described by SPTs with action  $\omega_{\pm}$.}\label{fig1}
\end{figure}

The problem becomes more interesting when we have two distinct $G$ invariant flows (labelled $\pm$) which each result in a trivially gapped phase. In this case there is an invariant (i.e.\ scheme independent) meaning to the difference in the SPTs resulting from the two flows.  In other words, the ratio of infrared actions
\begin{equation}\label{SPTacttrans}
    \frac{Z_{IR,+}[A]}{Z_{IR,-}[A]}=\exp\left(2\pi i \int_{X}\omega_{+-}(A)\right)~,
\end{equation}
is an invariant feature of the conformal field theory.\footnote{The fact that the CFT can form a transition between distinct SPTs also signals an anomaly in the coupling constant space of the theory \cite{Cordova:2019jnf, Cordova:2019uob, Hsin:2020cgg}. }  This setup has a natural interpretation as a second order phase transition between $G$-invariant trivially gapped phases: by deforming parameters the gap closes and the dynamics of the CFT can provide a transition to a new distinct trivially gapped phase (See Figure \ref{fig1}). Such SPT transitions have been previously explored in \cite{TSUI2015330, Tsui:2017ryj, 2017, TSUI2019114799, PhysRevB.100.165132, Lanzetta:2022lze, PhysRevLett.59.799, PhysRevB.93.165135, Nonne_2013, Roy:2015ars,Miyaji:2014mca, Cho2017UniversalES, Cho:2016xjw, Han:2017hdv, Cho:2015ega, Chen_2013}.

Of course, it is also possible to have a first order transition, with multiple gapped ground states that achieves a transition between SPTs. We describe an example of this in more detail in Appendix \ref{appendixfirstordertransitions}. Throughout the remainder of this work, we assume that the transition theory is second order with a unique vacuum and hence a conformal fixed point. A motivating question for the analysis to follow is thus, given a $G$ invariant conformal field theory with a pair of $G$ invariant flows to trivially gapped phases and a non-vanishing transition action $\omega_{+-}(A),$ can we strengthen the bound \eqref{cpos}?  It is intuitively clear that the answer to this question must be affirmative. The trivial CFT (with vanishing central charge $c$) cannot make a transition between distinct SPT phases, and thus the non-vanishing transition action $\omega_{+-}(A)$ necessitates local degrees of freedom with strictly positive $c$.  Our goal is to quantify this idea.

A simple example illustrating the ideas described above is given by the group $G\cong \mathbb{Z}_{2}$.  For bosonic theories there are no possible SPT transitions to consider. Indeed, such phases are classified by $H^{2}(G,U(1))$ which vanishes with $G\cong\mathbb{Z}_{2}$. However, for fermionic theories there is a unique possible non-trivial transition specified by the $\mathbb{Z}_{2}$-SPT defined by the non-trivial phase of the Kitaev chain \cite{Kitaev:2000nmw}. Formally this action may be written as:
\begin{equation}\label{SPTacttransex}
    \frac{Z_{IR,+}[A]}{Z_{IR,-}[A]}=\exp\left(\pi i \int_{X}q_{\rho}(A)\right)~,
\end{equation}
where $q_{\rho}(A)$ is a quadratic function of the $\mathbb{Z}_{2}$ background gauge field $A$, and $\rho$ is a fixed reference spin structure.\footnote{Alternatively, the quantity $q_{\rho}(A)$ in \eqref{SPTacttransex} is the Arf invariant of the spin structure $\rho+A$.} Famously, the $\mathbb{Z}_{2}$-SPT transition defined by \eqref{SPTacttransex} may be achieved through a massless Majorana fermion $\chi$ which is odd under the $\mathbb{Z}_{2}$ symmetry.  The mass deformation of this model deforms the Lagrangian by a term $m \chi^{2}$ where $m$ is a real parameter which may have either sign.  Both signs lead to trivially gapped physics preserving the $\mathbb{Z}_{2}$ symmetry, but differ by an SPT \eqref{SPTacttransex}:
\begin{equation}\label{flowex}
   \lim_{m\rightarrow \infty} \frac{Z_{\chi}[A,+m]}{Z_{\chi}[A,-m]}=\exp\left(\pi i \int_{X}q_{\rho}(A)\right)~.
\end{equation}
In fact, it is straightforward to see that the theory of the massless Majorana fermion is the CFT with smallest possible central charge $c$ that can achieve the transition \eqref{SPTacttransex} since all other non-trivial unitary CFTs have strictly greater central charge.

Below we will derive a general bound on $c$ for conformal field theories that can transition between $G$-SPTs for general group $G$.  For simplicity we focus on the case of bosonic theories where the SPTs may be classified by group cohomology classes:
\begin{equation}
    \omega_{+-}(A)\in H^{2}(G,U(1))~.
\end{equation}
Working with rational CFTs, those with a finite number of current algebra primaries, we derive a rigorous inequality relating the central charge, the SPT transition class and the spectrum of light operators in the CFT.  We subsequently illustrate this bound in several concrete examples of symmetry protected flows.

\subsection{Boundary Conformal Field Theory and Inflow}\label{inflowintro}

The main conceptual insight that enables our derivation is to relate the analysis of symmetry protected renormalization group flows to a problem in boundary conformal field theory \cite{Miyaji:2014mca, Cho2017UniversalES, Cho:2016xjw, Han:2017hdv, Cho:2015ega}.  Conformal boundary conditions arise naturally by activating the deformation \eqref{relopdef} in only half of spacetime:
\begin{equation}\label{halfdef}
     \delta S= \lambda \int_{x>0} dydx \ \mathcal{O}(y,x)~.
\end{equation}
Macroscopically, such a deformation results in a conformal interface between the original CFT (in the region $x\ll 0$) and the long distance limit of the relevant perturbation \eqref{relopdef} (in the region $x\gg 0$).  Such interfaces have been previously investigated in \cite{Gaiotto:2012np, Fredenhagen:2005an, Fredenhagen:2009tn, Brunner:2007ur, Brunner:2008fa, Cardy:2017ufe, Chang:2018iay}.  In general, a precise definition of the interface \eqref{halfdef} may also involve activating relevant operators along the interface locus.  In particular, in the special case where the bulk deformation \eqref{relopdef} results in a trivially gapped theory the half space deformation \eqref{halfdef} results, at long distances, in a conformal boundary condition for the initial CFT.

As a result of this correspondence, a conformal field theory which can make a transition between distinct SPT phases is endowed with two necessarily distinct conformally invariant boundary conditions which preserve the symmetry needed to define the SPTs.  Moreover, we may view the phase transition as occurring along a real spatial direction $x$ by activating the deformation in a strip: for $x\ll 0$ the CFT is deformed by the relevant perturbation $\mathcal{O}_{-}$, for $x\gg 0$ the CFT is deformed by the operator $\mathcal{O}_{+}$, and for a macroscopic region in the middle the CFT is undeformed. At long distances this results in the CFT defined on the strip geometry illustrated in Figure \ref{fig1} with distinct conformal boundary conditions arising from deformations. 

\begin{figure}[t]
    \centering
    \begin{subfigure}[h]{\textwidth}
        \centering
        %\includegraphics[width=0.475\linewidth]{DomainWall.pdf}%
        %\hfill
        \includegraphics[width=0.475\linewidth]{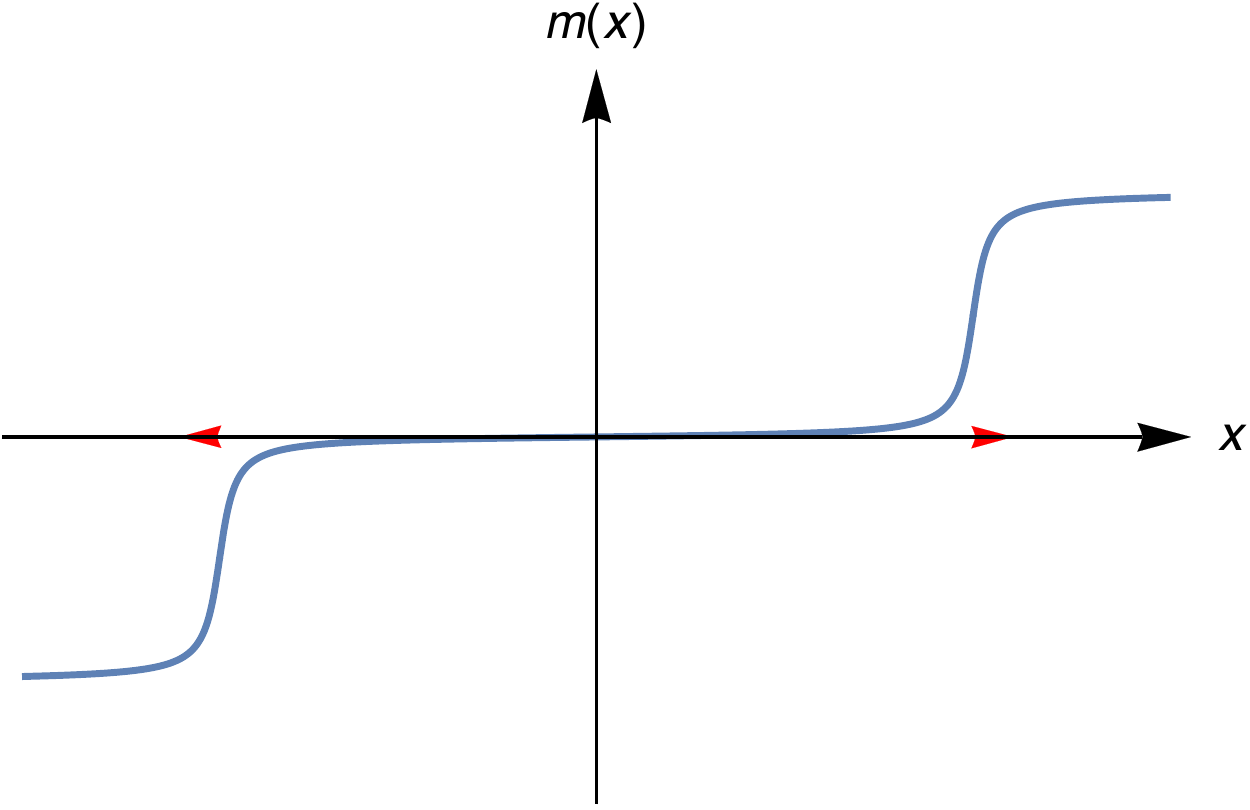}%
    \end{subfigure}
    \caption{The mass profile $m(x)$ of the deformed Majorana fermion. In the region where $m \approx 0$ we have a (fermionic) CFT and at the far left and far right we have different SPT phases. Renormalization group flow develops some conformal boundary conditions that interpolate in the IR between the regions with $|m|> 0$ and the interior where the fermion remains massless. } \label{fig2}
\end{figure}

For instance, in the fermionic $\mathbb{Z}_{2}$-SPT example discussed around \eqref{SPTacttransex}, the flows are triggered by mass deformations of the Majorana fermion $\chi$ with opposite signs (see \eqref{flowex}). So, in this example the boundary CFT setup just described can be envisioned by turning on different signs of the mass deformation $m \chi^{2}$ at the far left and far right, as depicted in Figure \ref{fig2}. In the IR, conformal boundary conditions are generated that transition between the trivially gapped phase of massive fermions and the massless fermions in the interior.

A conformal field theory defined on a spatial segment with boundary conditions defines an effective quantum mechanics.  Since both operator deformations preserve the $G$ symmetry so do the resulting boundary conditions and therefore the resulting quantum mechanics is $G$ invariant.  Moreover, the SPT transition class $\omega_{+-}$ defined in \eqref{SPTacttrans} has a simple direct interpretation as the anomaly in the $G$ symmetry of this quantum mechanics.  This can be understood from Figure \ref{fig1} from anomaly inflow: the bulk SPT regions to the far right and far left each contribute resulting in a total anomaly $\omega_{+-}(A)$. 

This class of anomalies, characterized by group cohomology, has a straightforward physical meaning: all states are in projective representations of $G$ characterized by the associated cohomology class. Recall that a projective representation  of a group $G$ is a group action on a vector space where multiplication in $G$ holds only up to phases:
\begin{equation}
    g_{1} ( g_{2} \vec{v}) =\exp(i\omega(g_{1},g_{2}))(g_{1}g_{2})\vec{v}~,
\end{equation}
and where the phase function $\omega$ cannot be removed by redefining the matrices $g_{i}.$  As is well known a projective representation of $G$ may equivalently be viewed as an ordinary representation of a larger group, $\widetilde{G}$ extending $G$ by additional central elements $Z$ which act as phases:
\begin{equation}
    1\rightarrow Z \rightarrow \widetilde{G}\rightarrow G \rightarrow 1~.
\end{equation}
A familiar example which occurs below is the case where $G\cong PSU(M) \cong SU(M)/\mathbb{Z}_{M},$ $\widetilde{G}\cong SU(M),$ and $Z\cong\mathbb{Z}_{M}.$  An important point to emphasize is that all states are in representations with a fixed action of the central elements $Z$.  Local operators in the quantum mechanics, all transform in ordinary representations of $G$ and hence cannot modify the action of $Z$.

One important consequence of this is the fact that, for non-zero $\omega_{+-}\in H^{2}(G,U(1)),$ every energy level must be degenerate.  Indeed, any non-trivial projective representation is  necessarily of dimension strictly larger than one.  In particular, for any CFT which can make a transition between distinct SPT states, the ground state degeneracy $d_{AB}$ on a strip with conformal boundary conditions $A$ and $B$ defined by the deforming operators is necessarily larger than unity:
\begin{equation}
    H^{2}(G,U(1))\ni\omega_{+-}\neq 0 \Longrightarrow d_{AB}>1~.
\end{equation}
Crucially, this fact allows us to quantify the abstract difference between the SPT phases, characterized by a cohomology class $\omega_{+-}$ in terms of the ground state degeneracy $d_{AB},$ a number.  Below when we present general bounds on the central charge $c$, the SPT difference will appear implicitly through the degeneracy $d_{AB}.$

\subsection{Constraints from Modular Invariance}

As outlined above, conformal field theories which can make transitions between distinct SPT phases are naturally equipped with conformal boundary conditions $A$ and $B$ and a ground state degeneracy $d_{AB}>1.$ To relate this data to the central charge of the bulk CFT, we invoke modular invariance.

As first described in the seminal work of Cardy \cite{Cardy:1989ir}, a conformally invariant boundary condition is related by a modular transformation to a state $|A\}$ in the standard Hilbert space of the CFT quantized on a circle. Consistency of the Hilbert space interpretation of $|A\}$ implies constraints on the state $|A\}$ which have been investigated in a variety of contexts including \cite{Cardy:1989ir, Cardy:1991tv,Cardy:2004hm,LEWELLEN1992654, Behrend:1999bn, Pradisi:1996yd, Fuchs:1997kt, Fuchs:2009zr}. In the special case of rational CFTs the constraints on boundary states are explicitly solvable and focusing on this class of examples enables us to derive a simple universal inequality relating the partition function of a rational CFT quantized on an interval with boundary conditions $A$ and $B$, and the partition function of the CFT quantized on a circle (i.e. without boundary conditions):
\begin{equation}\label{inttorineq}
    \mathrm{Tr}_{\mathcal{H}_{AB}}\left[e^{-\beta\left(L_{0}-\frac{c}{24}\right)}\right]\leq \mathcal{D}~\mathrm{Tr}_{\mathcal{H}_{S^{1}}}\left[e^{-2\beta\left(L_{0}-\frac{c}{12}\right)}\right]~,
\end{equation}
where above, $\mathcal{D}$ is the total quantum dimension.  

\begin{figure}[t]
        \centering
        \includegraphics[width=0.8\textwidth]{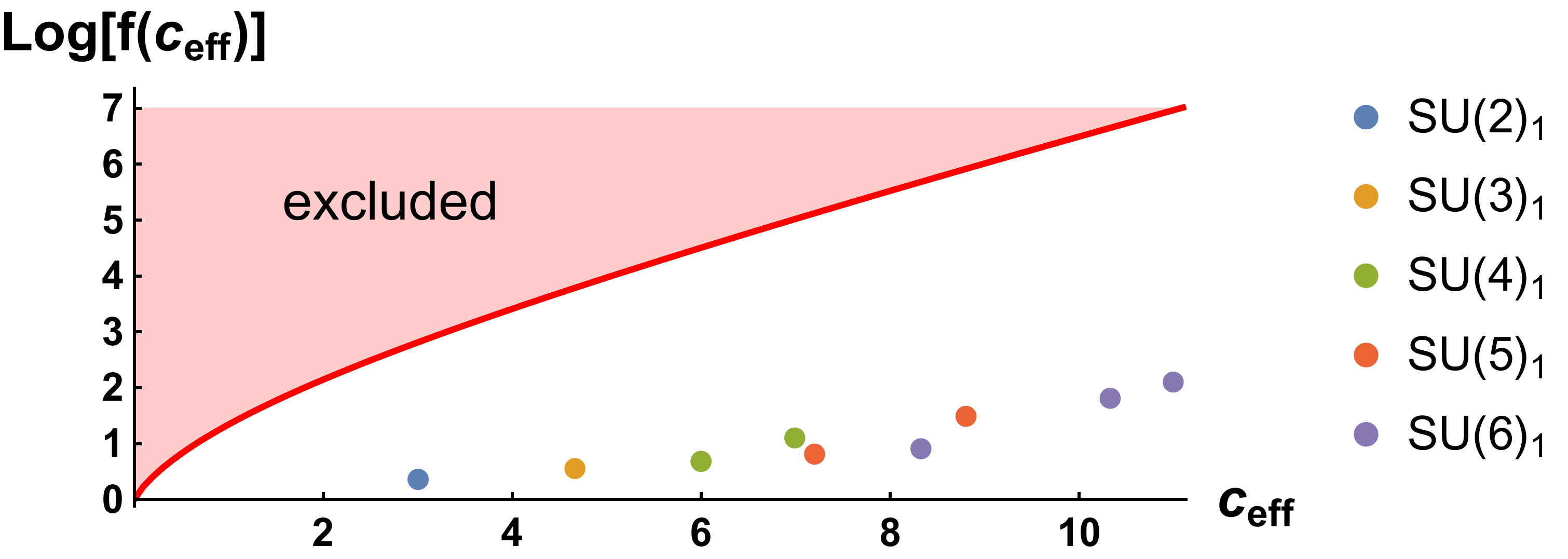} 
        \caption{The function $f(c_{\text{eff}})$ appearing in the inequality \eqref{boundintro}.  The interval ground state degeneracy $d_{AB}$, weighted by the light primaries $N$ and total quantum dimension $\mathcal{D}$, is constrained to lie below the curve.  The points show possible SPT transitions for $G\cong PSU(M)$ for various $M$ mediated by the WZW models $SU(M)_{1}.$}\label{figiplot}
\end{figure}

The inequality \eqref{inttorineq} gives a general constraint between the bulk CFT data, including the spectrum of local operators and central charge $c$ and the spectrum of the theory on the interval, including the ground state degeneracy $d_{AB}$. We can extract a sharp relationship between these data by generalizing the analysis of \cite{Hartman:2014oaa, Hellerman:2009bu} which enables us to estimate the torus partition function at finite temperature $\beta$ via the spectrum of light bulk operators.  Carefully taking into account the contributions of currents defining the extended chiral algebra yields a final bound:
\begin{equation}\label{boundintro}
   \frac{ d_{AB}}{\mathcal{D}N}\leq f(c_{\text{eff}})~,
\end{equation}
where above, $N \coloneqq N_{0A} + N_{B}$ and $c_{\text{eff}}$ are respectively the number of light bulk primary operators and the effective central charge, defined more precisely as: 
\begin{align}
     N_{B} = \# \text{Virasoro Primaries with} \ \Delta = &h + \overline{h} < \frac{c-1}{12}+\frac{1}{2\pi}~, \qquad h \neq 0~, \quad \text{and} \quad \overline{h} \neq 0~, \\[0.4cm]
    N_{0A} = \# \text{Virasoro Primaries with} \ \Delta = &h + \overline{h} < \frac{c-1}{12}+\frac{1}{2\pi} + g_{*}~, \quad h = 0~, \quad \text{or} \quad \overline{h} = 0~, \\[0.4cm]
     &c_{\text{eff}}\equiv c+8h_{AB,\text{min}}~,
\end{align}
where $h$ and $\overline{h}$ are the conformal weights of the Virasoro primary, $g_{*} \approx 0.0123...$ is a constant whose origin is explained below, and $h_{AB,\text{min}}$ the smallest scaling dimension (i.e.\ ground state energy) of states in the $AB$ interval sector. The function $f$ appearing in \eqref{boundintro} is defined below as the solution to an optimization problem.  A plot is shown in Figure \ref{figiplot}.  
An important comment is that, although we have motivated this inequality via renormalization  group flows, the result holds more generally for choice of boundary conditions in a rational diagonal conformal field theory.

We obtain a simplified version of the inequality \eqref{boundintro}, by examining the limit of large central charge $c_{\text{eff}}\rightarrow \infty$.  Assuming for simplicity that the number of light Virasoro primaries and total quantum dimension do not grow exponentially fast with $c_{\text{eff}}$
\begin{equation}\label{growth}
    \frac{\log{N}}{c_{\text{eff}}}\rightarrow 0~, \hspace{.2in}\frac{\log{\mathcal{D}}}{c_{\text{eff}}}\rightarrow 0~,
\end{equation}
we deduce that the dominant contribution to the inequality comes from the ground state degeneracy $d_{AB}$ and exponential growth in the function $f(c_{eff})$ implying parametrically
\begin{equation}   \label{asympbound} 
\log(d_{AB})\leq \frac{\pi}{8} c_{\text{eff}}~, \hspace{.2in}c_{\text{eff}}\rightarrow \infty~.
\end{equation}
A bound of this qualitative form, relating the logarithm of the ground state degeneracy and the central charge, was previously considered in \cite{2017} and conjectured to hold even at small $c$. By contrast our result, derived rigorously from modular invariance, only asymptotically takes this form under the assumptions highlighted above.  More generally, for small $c$ or even asymptotically when the assumptions \eqref{growth} are not satisfied, one must use the inequality \eqref{boundintro}.\footnote{\label{footcounter}In fact the bound conjectured in \cite{2017}, explicitly $\log_{2}(d_{AB})\leq c$, is falsified e.g.\ by considering $(G_{2})_{1}$ with Cardy boundary conditions associated respectively to the identity and the unique non-trivial integrable representation of dimension $7$.  Then,  $d_{AB}=7$ and $c=14/5$. By contrast, the bound \eqref{boundintro} is comfortably satisfied with $c_{eff}=6$ and $\log(f(c_{eff}))\approx 4.5.$ }

\subsection{Flows From WZW Models}

In section \ref{section4} below, we explore the ideas described above in the explicit example of SPT transitions for the group $G\cong PSU(M)\cong SU(M)/\mathbb{Z}_{M}$.  The possible SPTs for this symmetry are then classified by the cohomology group:
\begin{equation}
    H^{2}(PSU(M),U(1))\cong \mathbb{Z}_{M}~.
\end{equation}
When the transition class $\omega_{+-}$ takes a particular value $p$ mod $M$ in the group above, the resulting ground states transform in a projective representation which may be viewed as ordinary representations of $SU(M)$ defined by having $p \mod M$ boxes in the associated Young tableaux.   

A particular CFT with $G$ symmetry is the WZW model $SU(M)_{1}$. Using the ideas of \cite{Cardy:2017ufe}, we show how to achieve all possible SPT transitions for $G$ by renormalization group flows and describe the dictionary between boundary states and perturbing operators generating the flow.  This generalizes the closely related analysis of $SU(2)$ WZW models carried out in \cite{Chen_2013}.  (In Appendix \ref{Other-Simply-Laced-Algebras}, we similarly describe other flows for other WZW$_{1}$ CFTs based on simply laced Lie algebras.) Some examples of the data for these flows relative to the inequality \eqref{boundintro} are shown in Figure \ref{figiplot}.  In particular, by studying large rank examples we can realize boundary conditions such that:
\begin{equation}
    c_{eff}\sim \log(d_{AB}) \sim \log(N) \gg 1~.
\end{equation}
Thus, the asymptotic validity of the general bound \eqref{boundintro} is sensitive to the particular $\mathcal{O}(1)$ coefficients.  

Finally, in section \ref{section5} below we describe an example of a transition based on a discrete symmetry group $G\cong \mathbb{Z}_{2}\times \mathbb{Z}_{2}$.  Additional supplemental material is presented in the appendices.

\section{Boundary CFT and Renormalization Group Flows} \label{section2}

In this section we review boundary conformal field theory in rational CFTs following \cite{Cardy:1989ir}. (See also the review \cite{Andrei:2018die} and references therein for a recent discussion.) We also describe the map between relevant operators and conformal boundary conditions using the ansatz of \cite{Cardy:2017ufe}.

\subsection{Review of Boundary CFT}\label{bcftreviewsec}

Let us review the general framework of boundary conformal field theory.  Throughout this section we assume that the underlying bulk CFT is rational.  Thus, under the extended chiral algebra there are a finite number of primary operators labelled by an index $i$.  The Hilbert space on the plane is then decomposed into a sum of irreducible representations as $\mathcal{H}_{i} \otimes \mathcal{H}_{\overline{j}}$, and $\chi_{i}(\tau)$ and $\overline{\chi}_{\overline{j}}(\overline{\tau})$ are the characters of the respective representations.  The torus partition function is
\begin{equation} \label{partitionfunctionontheplane}
    Z_{T^{2}}(\tau,\overline{\tau}) = \sum_{i,\overline{j}} M_{i,\overline{j}} \chi_{i}(\tau)\overline{\chi}_{\overline{j}}(\overline{\tau})~,
\end{equation}
where the $M_{i,\overline{j}}$ are non-negative integers and the sum is finite.

\begin{figure}[t]
        \centering
        \includegraphics[width=0.55 \textwidth]{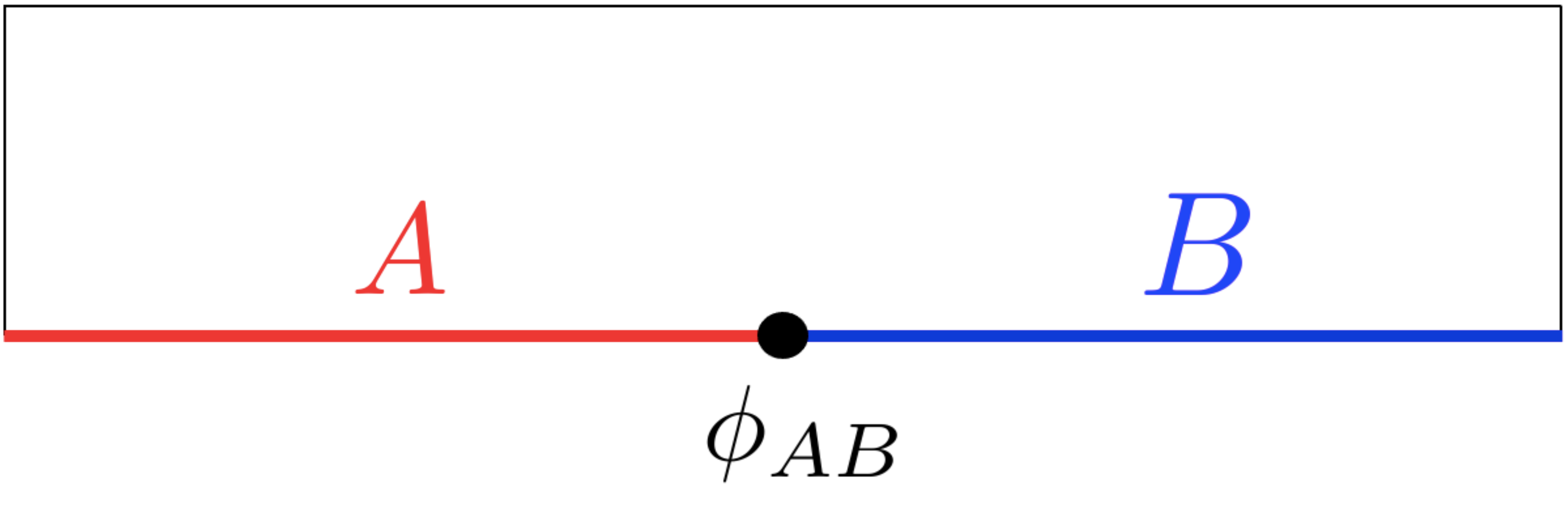} % second figure itself
        \caption{On the upper half-plane the conformal boundary conditions $A$ and $B$ appear on the two halves of the boundary (the real line), with a boundary-condition-changing-operator $\phi$ at the junction.}\label{fig4}
\end{figure}

In boundary conformal field theory we are concerned with placing the CFT on a spatial interval of finite extent.  At the ends of the interval are boundary conditions $A$ and $B$.  The boundary conditions are assumed to preserve conformal symmetry (see below).  The primary quantity of interest is then the cylinder partition function which counts the states on the interval weighted by their energy: 
\begin{equation} \label{openchannelpartition1}
    Z_{AB}(\beta) = \mathrm{Tr}_{\mathcal{H}_{AB}} \exp(-\beta H)~,
\end{equation}
where $\mathcal{H}_{AB}$ denotes the Hilbert space of the CFT with the indicated boundary conditions.  The state operator correspondence allows us to relate the states in $\mathcal{H}_{AB}$ to boundary operators.  Specifically, an infinite strip is conformal to the upper half-plane where the boundary of the strip is mapped to the real line $\mathrm{Im}(z) = 0$. The data of a state in $|\phi\rangle \in \mathcal{H}_{AB}$ is then mapped to a boundary operator $\phi$ that forms a junction between the boundary conditions $A$ and $B$ (See Figure \ref{fig4}).

The conformal transformation discussed above allows us to relate the energy of states in the cylinder, measured by the Hamiltonian $H$ in \eqref{openchannelpartition1}, to the scaling dimension of the associated boundary operators, measured by the Virasoro generator $L_{0}$ as:
\begin{equation}
    H=L_{0}-\frac{c}{24}~,
\end{equation}
where $c=\bar{c}$ is the central charge.\footnote{We assume throughout our analysis that the CFT is non-chiral ($c=\bar{c}$) so that conformal boundary conditions exist.}. In this case it is customary to express the partition function as 
\begin{equation} \label{openchannelpartition}
    Z_{AB}(q) = \mathrm{Tr}_{\mathcal{H}_{AB}} (q^{L_{0}-c/24})~, \quad q \equiv e^{-\beta}~.
\end{equation}

Of particular importance to us is the ground state in the Hilbert space $\mathcal{H}_{AB}$.  If the left and right boundary conditions are equal, then the identity operator, of scaling dimension zero, is an allowed boundary operator and hence corresponds under the state operator map to the ground state.  A conformal boundary condition $A$ is called \emph{elementary} when the degeneracy of the unit operator in $\mathcal{H}_{AA}$ is one.\footnote{A non-elementary boundary condition which has multiple boundary unit operators may be viewed as an interface between the bulk CFT and a non-trivial topological quantum field theory. In this case, the multiple boundary unit operators correspond to the local operators in the topological theory. See e.g.\ \cite{Chang:2018iay} for related discussion. }    If the boundary conditions $A$ and $B$ are distinct, then the lowest dimension primary boundary operator in general has a conformal weight $h_{AB,\text{min}}>0.$  The large $\beta$ behavior of the partition function $Z_{AB}(\beta)$ is then dominated by this operator so that 
\begin{equation}\label{intgroundstate}
    Z_{AB}(\beta)=d_{AB}\exp\left[-\beta\left(h_{AB,\text{min}}-\frac{c}{24}\right)\right]+\cdots~,
\end{equation}
where the omitted terms are subleading for large $\beta$.  The positive integer $d_{AB}$ is then the ground state degeneracy.  As discussed in section \ref{inflowintro} above, if the bulk conformal field theory interpolates between distinct SPT phases the degeneracy $d_{AB}$ for the associated RG boundary conditions is necessarily larger than one.

Modular invariance provides powerful constraints on conformal boundary conditions and the cylinder partition function.   By an $S$-transformation, we may reinterpret the partition function $Z_{AB}$ as a transition amplitude between two associated \emph{boundary states} $|A\}$ and $|B\}$ in the Hilbert space $\mathcal{H}_{S^{1}}$ of the bulk CFT quantized on a circle:  
\begin{equation} \label{closedchannelpartition}
    Z_{AB} = \{ \Theta B | \tilde{q}^{\frac{1}{2}(L_{0} + \overline{L}_{0}-(c + \overline{c})/24)} | A \}~, \quad \tilde{q} = e^{ - 4 \pi^{2} / \beta}~,
\end{equation}
where above the (anti-unitary) bulk CPT operator $\Theta$ appears to account for the fact that the the two boundary components of the cylinder have opposite orientation.

Boundary states are constrained by symmetry considerations. They carry the information that conformal symmetry is preserved on the theory with a boundary.  Working on the upper half-plane with boundary on the real line the boundary restrictions of the stress tensor are related as
\begin{equation}
    T(z)\big|_{\mathrm{Im} z = 0} = \overline{T}(\overline{z})\big|_{\mathrm{Im} z = 0}~.
\end{equation}
For the associated boundary state, this implies the gluing conditions:
\begin{equation}
    (L_{n}-\overline{L}_{-n})|A\}  =  0~, \quad \forall n~, \label{conformalgluing}
\end{equation}
where $L_{n},\,\overline{L}_{-n}$ are the modes of the stress-energy tensor. In the following, when we call a boundary state conformal, we do it to emphasize that this condition is satisfied.  

If the original bulk theory contains an extended symmetry algebra $\mathcal{W} \times \mathcal{W}$ beyond the Virasoro algebra, we can also consider the possibility of studying boundary conditions that preserve (half of) such an extended symmetry. For the associated boundary state this results in additional gluing conditions
\begin{equation}\label{extendedgluing}
    \big( W^{(s)}_{n}-(-1)^{h^{(s)}}\Omega(\overline{W}^{(s)}_{-n}) \big) |A\} = 0~, \quad \forall n~,  
\end{equation}
where $W^{(s)}_{n}$ are the modes for the corresponding symmetry generator (the label $(s)$ labels distinct currents).  We stress that even if an extended symmetry exists in the bulk CFT, there may be boundary states that only satisfy \eqref{conformalgluing} and not \eqref{extendedgluing}, in which case no extended continuous symmetry is present in the theory with a boundary. In equation \eqref{extendedgluing} we have also introduced $\Omega: \mathcal{W} \rightarrow \mathcal{W}$ a local automorphism of the chiral symmetry algebra, which leaves the energy-momentum tensor fixed $\Omega T = T$. This means $\Omega$ commutes with the mode expansion of the chiral generators $W^{(s)}(z)$, and that $\Omega$ is compatible with operations in the $W$-algebra such as taking commutators or derivatives. 
The automorphism $\Omega$ induces an action of elements $w \in \mathcal{W}$ on $\mathcal{H}_{i}$ via $\pi^{\Omega}_{i}(w)|h\rangle = \pi_{i}{(\Omega w)}|h\rangle$, where $\pi_{i}(\omega)$ denotes the action of $w$ over $|h\rangle$ for trivial $\Omega$. Then, $\mathcal{H}_{i}$ equipped with this new action is isomorphic to some $\mathcal{H}_{\Omega(i)}$. The isomorphism is implemented by a unitary operator:
\begin{equation} \label{VOmega}
    V_{\Omega}: \mathcal{H}_{\Omega{(i)}} \rightarrow \mathcal{H}_{i}.
\end{equation}

To proceed further, we now follow \cite{Ishibashi:1988kg} and solve the gluing conditions.  Assuming the automorphism $\Omega$ is trivial, for each representation $[\phi_{i}]$ of the chiral symmetry algebra $\mathcal{W}$ a solution $| i \rrangle$ of the gluing conditions can be found which takes the form:
\begin{equation} \label{Ishibashi-States}
    | i \rrangle = \sum_{N=0}^{\infty} | i,N \rangle \otimes  U \overline{| i,N \rangle},
\end{equation}
where $|i, N \rangle$, $N \in \mathbb{N}^{+}$ denotes an orthonormal basis of $\mathcal{H}_{i}$ and $U$ is an antiunitary operator which satisfies $U \overline{W}_{n} = (-1)^{h_{W}}\overline{W}_{n}U$. Such a state $ | i \rrangle$ is known as the Ishibashi state associated to $i$.\footnote{We can write Ishibashi states for non-trivial $\Omega$ by dressing the Ishibashi states for trivial $\Omega$ with the operator appearing in \eqref{VOmega}: $| i \rrangle_{\Omega} = (\mathbbm{1} \otimes V_{\Omega}) | i \rrangle $.} Ishibashi states are further characterized by their overlap regularized by an insertion of $q^{L_{0}-c/24}$:
\begin{equation}
    \llangle i | q^{L_{0}-c/24} |j \rrangle = \delta_{j,i} \, \chi_{i}(q).
\end{equation}

One can gain intuition about these states by naively taking a Ishibashi state to be a boundary state. Then Virasoro scalar states of the form $|i, N\rangle \otimes U \overline{| i,N \rangle}$ would have a non-zero overlap with the corresponding Ishibashi state only:
\begin{equation} \label{Ishibashi-overlap-with-Virasoroscalar}
    | \llangle j  |i, N\rangle \otimes U \overline{| i,N \rangle} | = \delta_{i,j}~.
\end{equation}
By a conformal transformation, one can reinterpret this inner-product as a one-point function of the bulk operator on the disc with the given boundary condition.  Thus, an Ishibashi state is a solution to the gluing conditions which, if interpreted naively as a boundary state, gives expectation values only to the states descending from a single primary and its conjugate.

Of course, the Ishibashi states are only building blocks which ensure the preservation of conformal and possibly extended symmetries.  True boundary states are obtained as linear combinations:
\begin{equation} \label{bdystates}
    | A \} = \sum_{i} \mathcal{B}_{A}^{\ i} \, |i\rrangle~.
\end{equation}
The coefficients in the expansion above are constrained by demanding the consistency of the boundary Hilbert space interpretation.  Evaluating the partition function \eqref{closedchannelpartition} using the states \eqref{bdystates} gives:
\begin{equation}
   Z_{AB}(q)= \{ \Theta B | \tilde{q}^{\frac{1}{2}(L_{0} + \overline{L}_{0}-c/12)} | A \} = \sum_{j} \mathcal{B}^{\ j^{+}}_{B} \mathcal{B}^{\ j}_{A} \chi_{j}(\tilde{q})~,
\end{equation}
which is an expression written in terms of characters evaluated at $\tilde{q}$, and where $j^{+}$ stands for the representation conjugate to $j$.  Using the fact that our underlying CFT is rational, we can reexpress the above in terms of characters evaluated at $q$ using the finite dimensional modular $S$-matrix:
\begin{equation}
    \chi_{j}(\tilde{q}) = \sum_{i} S_{ji} \chi_{i}(q)~.
\end{equation}
Hence we obtain:
\begin{equation}
    Z_{AB}(q)=\sum_{j,i} \mathcal{B}^{\ j^{+}}_{B} \mathcal{B}^{\ j}_{A} S_{ji} \, \chi_{i}(q)~.
\end{equation}
By definition, each character $\chi_{i}(q)$ has a consistent state counting interpretation.  Thus, the entire Hilbert space $\mathcal{H}_{AB}$ will also have such an interpretation provided that the coefficient of each character is a non-negative integer.  Therefore we require the combination
\begin{equation}\label{littlendef}
 n_{AB}^{i}\equiv   \sum_{j}\mathcal{B}^{\ j^{+}}_{B} \mathcal{B}^{\ j}_{A} S_{ji}~,
\end{equation}
to obey the integrality constraints:
\begin{equation}\label{Cardycondition}
    n_{AB}^{i} \in \mathbb{Z}~, \hspace{.2in}n_{AB}^{i}\geq0~, \hspace{.2in}n_{A^{+}A}^{0}=1~,
\end{equation}
where the last inequality above enforces the condition that the boundary condition is elementary and hence supports a unique boundary unit operator.  The corresponding partition function on the cylinder then takes the form
\begin{equation} \label{openchannelpartition}
    Z_{AB} = \mathrm{Tr}_{\mathcal{H}_{AB}} (q^{L_{0}-c/24}) = \sum_{i} n^{i}_{AB} \, \chi_{i}(q)~,
\end{equation}
so that $n_{AB}^{i}$ controls the degeneracy of the $i$-th character in the Hilbert space $\mathcal{H}_{AB}.$

Before discussing solutions to \eqref{Cardycondition} we note in passing two significant features:
\begin{itemize}
    \item The $\mathcal{B}^{\ i}_{A}$ coefficients are generically non-zero for all $i$, so in general the inner products of a Virasoro scalar state $|i, N\rangle \otimes U \overline{| i,N \rangle}$ with any boundary state $|A\}$ is non-vanishing.  Comparing with the discussion below \eqref{Ishibashi-overlap-with-Virasoroscalar} we see that this means that in general all such scalar operators acquire disc expectation values in the presence of a boundary.  

    \item The contribution of the identity Ishibashi state is particularly interesting.  Expanding $|A\}$ as
    \begin{equation}
    | A \} = g_{A} |0\rrangle + \sum_{i \neq 0} \mathcal{B}_{A}^{i} | i \rrangle~,
\end{equation}
    the quantity $g_{A}$ is known as the $g$-function \cite{PhysRevLett.67.161} which controls the high temperature asymptotics of the interval partition function :
    \begin{equation}
  \lim_{\beta\rightarrow 0}  \log\left(Z_{AA}(\beta)\right) = \frac{\pi c}{6} \frac{L}{\beta} + 2 \ln{(g_{A})} + \cdots~,
\end{equation}
    where above $L$ is the length of the interval and the neglected terms are subleading as $\beta$ tends to zero. The $g$-function may be used to generalize the $c$-theorem to the context of boundary CFT \cite{Friedan:2003yc, PhysRevB.48.7297, Casini:2016fgb}. See also \cite{Friedan:2012jk, Friedan:2013bha, Collier:2021ngi} for more recent results on constraints on $g$ using bootstrap techniques.
\end{itemize}

Returning to the discussion of \eqref{Cardycondition}, a concrete solution for trivial $\Omega$ and for diagonal rational CFTs was originally worked out by Cardy in \cite{Cardy:1989ir}.  The key step is to make use of the Verlinde formula \cite{VERLINDE1988360}:
\begin{equation} \label{VerlindeFormula}
    N_{ij}^{k} = \sum_{m} \frac{S_{im}S_{jm} \overline{S}_{k m}}{S_{0m}}~,
\end{equation}
which relates the fusion coefficients $N_{ij}^{k}$ ($[\phi_{i}] \times [\phi_{j}] = \sum_{k} N_{ij}^{k} [\phi_{k}]$) and the unitary $S$ matrix.\footnote{Here we use that $\overline{S}_{ij} = S_{i j^{+}} = S_{i^{+} j}$. Additionally, the conjugate representation satisfies $N^{0}_{ij} = \delta_{j \, i^{+}}$.} In this solution, the boundary states $|a\}$ have the same labels as the irreducible representations of $\mathcal{W}$, and the explicit expression for these ``Cardy states'' expanded in terms of the Ishibashi states is
\begin{equation} \label{cardystates}
    |a\} = \sum_{i} \frac{S_{ai}}{\sqrt{S_{0i}}} |i\rrangle~.
\end{equation}
To see that this gives a well defined Hilbert space we evaluate equation \eqref{littlendef} to obtain
\begin{equation}
    n_{ab}^{i}= \sum_{j} \frac{S_{bj^{+}}S_{aj}S_{ji}}{S_{0j}} = \sum_{j} \frac{S_{aj}S_{ij}\overline{S}_{jb}}{S_{0j}} = N_{ai}^{b} = N_{a^{+}b}^{i}~.
\end{equation}
The constraints specified in equation \eqref{Cardycondition} are then satisfied, because the fusion coefficients are non-negative integers with 
$N^{0}_{a^{+}a}=1.$

To summarize, for a diagonal rational CFT in the spatial interval there are boundary conditions $|a\}$ preserving (half of) the extended chiral algebra.  Such boundary conditions are in one-to-one correspondence with the primaries of the chiral algebra and the interval partition function takes the form:
\begin{equation} \label{Cardy-OpenPartition}
    Z_{ab}(q) = \sum_{i} N^{i}_{a^{+} b} \, \chi_{i}(q)~,
\end{equation}
with $N^{i}_{a b}$ the corresponding fusion coefficients of the rational CFT. In general below, we will assume a rational but not necessarily diagonal CFT, although we often remark about the diagonal case where the above Cardy states are available.

\subsection{Boundary Conditions from Renormalization Group Flows} \label{subsection2.2}

As discussed in section \ref{inflowintro} we are often interested in the dictionary between renormalization group flows into trivially gapped phases, and conformal boundary conditions of the CFT (see Figure \ref{fig1}).  Here we summarize a ansatz for this correspondence developed in \cite{Cardy:2017ufe}. We apply this to specific examples in section \ref{subsubsection4.2.3}.

As before, we consider a CFT deformed by a set of relevant perturbations leading to a deformed Hamiltonian:
\begin{equation} \label{flowed-hamiltonian}
    H = H_{\mathrm{CFT}} + \sum_{j}  \int  dx  \ \lambda_{j}\mathcal{O}^{j} ~,
\end{equation}
where the $\mathcal{O}^{j}$ are relevant operators. We suppose that the deformation leads to a trivially gapped phase at long distances, i.e. in the formal limit $\lambda_{j}\rightarrow \infty$.  Our goal is then to determine which boundary state arises if the above deformation is activated in half of space.  Following \cite{Cardy:2017ufe}, it is natural to expect that the resulting boundary state $|A\}$ will be the one which minimizes the expectation value of the energy
\begin{equation}
    E_{A} = \frac{\{ A | H |A\}}{\{A|A\}}~.
\end{equation}
As it stands this does not quite form a satisfactory variational ansatz, since the boundary states do not have finite norm. To remedy this we can take the parameters $\lambda_{j}$ to be large but still finite.  In this case we expect the boundary state above to be perturbed by irrelevant operators. One such boundary irrelevant operator which is universally present is the bulk energy-momentum tensor.  Since the energy momentum tensor is neutral under all global symmetries this boundary operator is present in any symmetry preserving flow.  Allowing such a deformation means regulating the states as
\begin{equation}\label{changereg}
    |A\} \longrightarrow e^{-t_{A} H_{\mathrm{CFT}}} |A\}~,
\end{equation}
where $t_{A}$ is viewed as a function of $\lambda_{i}$, and tends to zero as $\lambda_{i}$ tends to infinity.  More generally, one could expect other symmetry preserving irrelevant operators to appear in the boundary deformation, but frequently the energy-momentum tensor gives the leading contribution.\footnote{For example in our applications in section \ref{subsubsection4.2.3} to WZW models, there are no lighter symmetry preserving operators so this ansatz applies.}

Assuming for simplicity that the regulation in \eqref{changereg} is sufficient, the resulting expression for the energy is then:
\begin{equation}
    E_{A} = \frac{\{ A | e^{-t_{A} H_{\mathrm{CFT}}} He^{-t_{A} H_{\mathrm{CFT}}} |A\}}{\{A|e^{-t_{A} H_{\mathrm{CFT}}} e^{-t_{A} H_{\mathrm{CFT}}}|A\}}~.
\end{equation}
We now minimize each $E_{A}$ with respect to $t_{A}$ then select among all boundary state labels $A$, the one with minimal energy.  The resulting global minimum is then the boundary state $|A\}$ arising from the flow \eqref{flowed-hamiltonian} driven by the given relevant operators.\footnote{It may happen that the variational algorithm above produces multiple degenerate boundary states realizing the same global minimum of the energy. In that case it is tempting to conjecture that the IR is not a trivially gapped phase, but rather yields a non-trivial TQFT at long distances. In this case the associated ``boundary state" described by such a flow is non-elementary should be the sum over the minima of the energy.}

The variational procedure outlined above was carried out in \cite{Cardy:2017ufe} using results from \cite{Cardy:1991tv} for case of rational diagonal CFTs where the space of boundary states is taken to be the Cardy states defined in \eqref{cardystates} above. One finds that the energies (viewed as a function of $t$ and $\lambda$, with $t$ assumed small) take the form:
\begin{equation} \label{Cardy-Energies}
    E_{a} = \frac{\pi c}{96 \, t_{a}^{2}} + \sum_{j \neq 0} \frac{S_{aj}}{S_{0a}} \frac{\lambda_{j}}{( t_{a})^{\Delta_{j}}}~,
\end{equation}
where above, we have rescaled the parameters $\lambda_{j}$ by a positive coefficient.  After minimizing each $E_{a}$ with respect to $t_{a}$ and choosing the $a$ that gives the global minimum, this procedure gives the candidate boundary condition. Thus, we see that whenever this variational analysis is valid the boundary state is determined solely by the modular data.

Finally, it is instructive to study the case of a single relevant deformation (one non-zero $\lambda_{j}$) and its conjugate in this framework. In this case \eqref{Cardy-Energies} reads
\begin{equation}
    E_{a} = \frac{\pi c}{96 t_{a}^{2}} + \frac{1}{t_{a}^{\Delta_{j}}}\frac{(S_{aj} \lambda_{j} + S_{aj^{+}}\lambda_{j^{+}})}{ S_{0a}}~.
\end{equation}
The energy $E_{a}$ appearing above is minimized when $t_{a}=t^{*}_{a}$ given by:\footnote{Notice that when $(S_{aj} \lambda_{j} + S_{aj^{+}}\lambda_{j^{+}})$ is positive for given $a$ there is no solution for a minima of $E_{a}$ at finite $t_{a}$. In applying this algorithm, we thus assume that for at least one label $a$ the coefficient $(S_{aj} \lambda_{j} + S_{aj^{+}}\lambda_{j^{+}})$ is negative, so that the global minima among all the $E_{a}$ is negative. Then, we obtain a solution for the global minima such that $t^{*}_{a}\rightarrow 0$ as $\lambda_{j}\rightarrow \infty$ and Cardy's variational problem has a well-defined solution. The critical point \eqref{criticalta} is therefore meaningful only when the corresponding coefficient $(S_{aj} \lambda_{j} + S_{aj^{+}}\lambda_{j^{+}})$ is negative.}
\begin{equation} \label{criticalta}
t^{*}_{a}  = \bigg( - \frac{\pi c S_{0a}}{48 \Delta_{j} (S_{aj} \lambda_{j} + S_{aj^{+}}\lambda_{j^{+}})} \bigg)^{\frac{1}{2-\Delta_{j}}} ~.
 \end{equation}
 Note that as expected as $\lambda_{j}\rightarrow \infty,$ the critical point $t^{*}_{a}\rightarrow 0.$ The resulting energies then take the form:
 \begin{equation}\label{Cardy-energies}
     E_{a}\Big|_{t^{*}_{a}}  = - \Big( \frac{\pi c}{96} \Big)^{-\frac{\Delta_{j}}{2-\Delta_{j}}} \Bigg[   \bigg( \frac{\Delta_{j}}{2} \bigg)^{\frac{\Delta_{j}}{2-\Delta_{j}}} -\bigg( \frac{\Delta_{j}}{2} \bigg)^{\frac{2}{2-\Delta_{j}}} \Bigg] \Bigg[- \frac{(S_{aj}\lambda_{j} + S_{aj^{+}}\lambda_{j^{+}})}{ S_{0a}}\Bigg]^{\frac{2}{2-\Delta_{j}}}~.
 \end{equation}

\section{Bounds on Partition Functions and the Central Charge} \label{section3}

In this section we derive bounds on the interval partition function and ground state degeneracy in terms of the bulk torus partition function and central charge. In section \ref{bulkbounds}, we derive a bound on the torus partition function in terms of the spectrum of light operators.  This analysis is a straightforward generalization of \cite{Hellerman:2009bu, Hartman:2014oaa}  to allow for the presence of additional holomorphic operators (applicable for instance to rational CFTs).  Next in section \ref{secintbound} we derive bounds on the interval partition function in terms of the bulk torus partition function.  Finally, in section \ref{groundstatedeg} we combine these results to deduce a bound on the ground state degeneracy on the interval in terms of bulk CFT data.  Throughout this section we take $c=\bar{c}>1$.

\subsection{Bounds on Partition Functions in Theories with Extended Chiral Algebras}\label{bulkbounds}

Consider the torus partition function $Z_{T^{2}}(\tau, \bar{\tau})$.  In general, we may split this into a sum of three terms as:
\begin{equation}\label{torustau}
    Z_{T^{2}}(\tau, \overline{\tau}) = Z_{00}(\tau, \overline{\tau}) + \sum_{A} Z_{0A}(\tau, \overline{\tau}) + \sum_{B}Z_{B}(\tau, \overline{\tau})~,
\end{equation}
where $Z_{00}$ is the contribution from the identity operator and its Virasoro descendants, $Z_{0A}$ is the contribution from holomorphic or antiholomorphic operators and their Virasoro descendants, and $Z_{B}$ is the contribution from all operators with Virasoro primaries of weight $(h,\overline{h})$ with both $h$ and $\overline{h}$ nonzero.

Our task is now to extract the contribution of the Virasoro descendants and express the above in terms of the primaries.  To this end, we recall that when $c > 1$ there are no Virasoro null vectors and descendants (from a chiral half of the algebra) may be arranged as
\begin{equation}
    L_{-n_{1}} L_{-n_{2}} \cdots L_{-n_{k}} |h\rangle~, \hspace{.2in} n_{1} \geq n_{2} \geq \cdots \geq n_{k}>0~.
\end{equation}
When $h=0,$ we have the further constraint that $n_{k}>1$.  We can write then, for instance:
\begin{equation}\label{Z00tau}
    Z_{00}(\tau, \overline{\tau}) = q^{-c/24} \overline{q}^{-c/24} \prod_{m=2}^{\infty}(1-q^{m})^{-1} \prod_{n=2}^{\infty}(1-\overline{q}^{n})^{-1} = \frac{q^{-\frac{(c-1)}{24}} \overline{q}^{-\frac{(c-1)}{24}}}{|\eta(\tau)|^{2}}(1-q)(1-\overline{q})~,
\end{equation}
where, as is standard, we have introduced $q$ and the Dedekind $\eta$-function
\begin{equation}
    q = e^{2 \pi i \tau}~, \hspace{.2in}\eta(\tau) = q^{1/24} \prod_{n=1}^{\infty}(1-q^{n})~.
\end{equation}
Below we are particularly interested in the special case where $\tau = - \overline{\tau} = i \beta / 2 \pi$ with $\beta$ real so that $q = \overline{q} = e^{-\beta}$.  In terms of $\beta$ \eqref{Z00tau} becomes
\begin{equation}\label{z0beta}
    Z_{00}(\beta) = \frac{e^{\beta\frac{(c - 1)}{12}} }{|\eta(\frac{i\beta}{2\pi})|^{2}}(1-e^{-\beta})^{2}~.
\end{equation}
Similarly, we have that:
\begin{equation}
    \sum_{A} Z_{0A}(\beta) = \frac{e^{\beta\frac{(c - 1)}{12}} }{|\eta(\frac{i\beta}{2\pi})|^{2}}(1-e^{-\beta}) \sum_{A} e^{-\beta \Delta_{A}}~,
\end{equation}
where $\Delta_{A}$ are the scaling dimensions of Virasoro primaries with the identity on either the holomorphic or antiholomorphic side but not both simultaneously. Meanwhile, for $Z_{B}(\beta)$ we can write
\begin{equation}\label{zbbeta}
    \sum_{B} Z_{B}(\beta) = \frac{e^{\beta\frac{(c - 1)}{12}} }{|\eta(\frac{i\beta}{2\pi})|^{2}} \sum_{B} e^{-\beta \Delta_{B}}~,
\end{equation}
where $\Delta_{B}$ are the scaling dimensions of Virasoro primaries with both $h$ and $\bar{h}$ non-zero. Collecting the contributions \eqref{z0beta}-\eqref{zbbeta} the full torus partition function \eqref{torustau} takes the form:
\begin{equation}
    Z_{T^{2}}(\beta) = \frac{e^{\beta\frac{(c - 1)}{12}} }{|\eta(\frac{i\beta}{2\pi})|^{2}} \Big[ (1-e^{-\beta})^{2} + (1-e^{-\beta})\sum_{A} e^{-\beta \Delta_{A}} + \sum_{B} e^{-\beta \Delta_{B}}  \Big]~.
\end{equation}

To bound this partition function, we follow the logic of \cite{Hartman:2014oaa}. Define the \emph{heavy} contribution, $Z_{H}(\beta)$ as the contribution to the partition function from operators of large dimension
\begin{equation}\label{zheavy}
    Z_{H}(\beta) \coloneqq \frac{e^{\beta\frac{(c - 1)}{12}}}{|\eta(\frac{i\beta}{2\pi})|^{2}} \Big[ (1-e^{-\beta})\sum_{\Delta_{A} \geq  \Delta_{AH}} e^{-\beta \Delta_{A} } + \sum_{\Delta_{B} \geq \Delta_{BH}} e^{-\beta \Delta_{B}} \Big]~,
\end{equation}
where $\Delta_{AH}$ and $\Delta_{BH}$ are thresholds, to be specified below, separating the heavy primaries from the light primaries. The \emph{light} piece of the torus partition function, $Z_{L}(\beta),$ is then defined such that $Z_{T^{2}}(\beta) = Z_{L}(\beta) + Z_{H}(\beta)$. 

\begin{figure}
        \centering
        \includegraphics[width=0.6\textwidth]{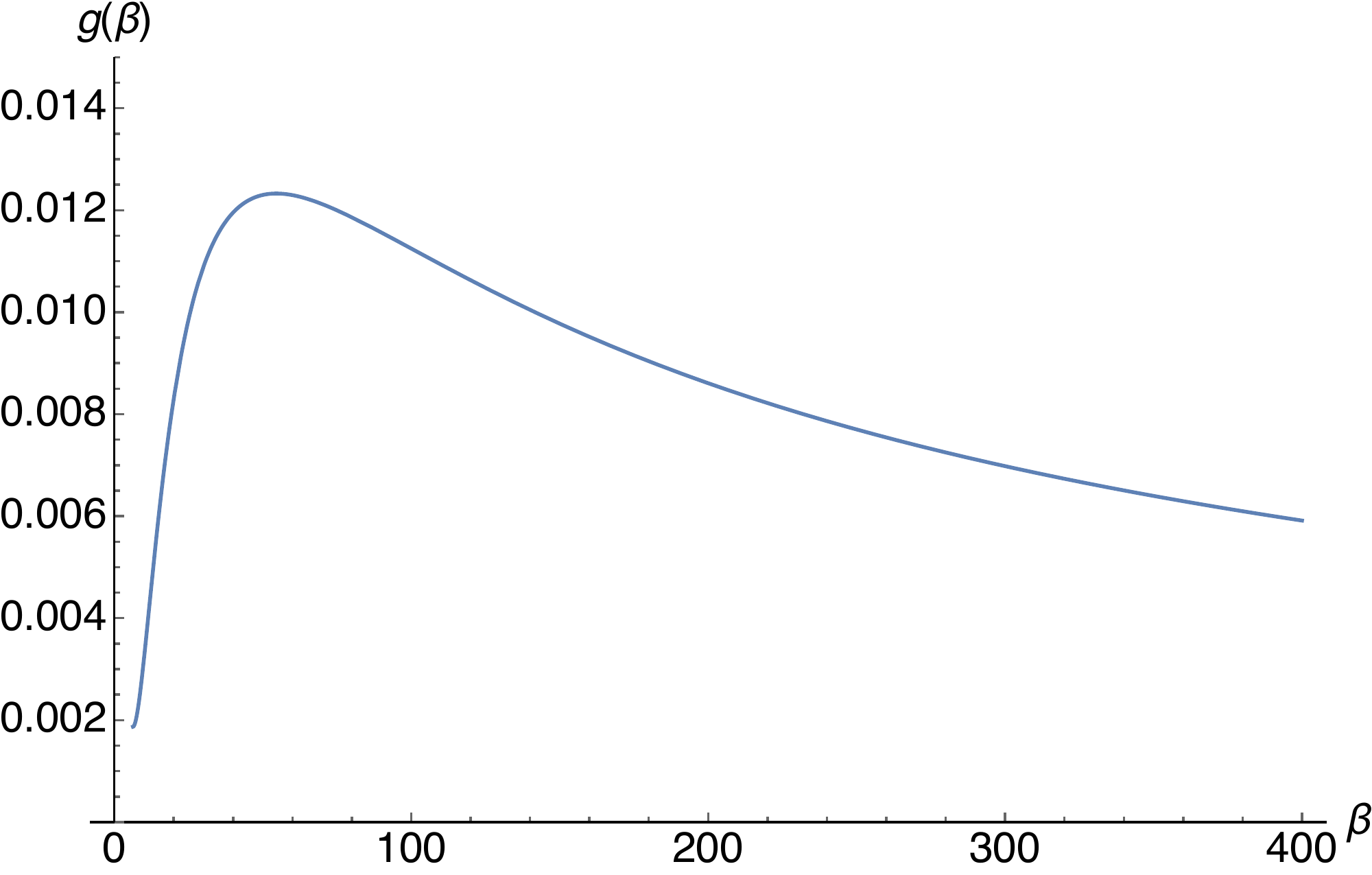} % second figure itself
        \caption{The function $g(\beta)$ defined in \eqref{DeltaAH-DeltabHoffset}. The function approaches $0$ as $\beta \to \infty$. The inequality \eqref{preineq} holds for all $\beta>2\pi$ provided we take $\Delta_{AH} - \Delta_{BH}$ larger than the maximum of $g(\beta)$.}\label{gbeta}
\end{figure}

Proceeding as in \cite{Hartman:2014oaa}, we wish to compare the contributions of heavy operators for the inverse temperatures $\beta$ and $\beta'\equiv 4\pi^{2}/\beta$ which are related by modular transformation.  Assuming $\beta>2\pi$ we notice that
\begin{equation}
     \sum_{\Delta_{B} \geq \Delta_{BH}} e^{-\beta \Delta_{B}} \leq e^{(\beta' - \beta) \Delta_{BH}} \sum_{\Delta_{B} \geq \Delta_{BH}} e^{-\beta' \Delta_{B}} \label{bound1}~,
\end{equation}
and similarly
\begin{equation}
    (1-e^{-\beta})\sum_{\Delta_{A} \geq  \Delta_{AH}} e^{-\beta \Delta_{A} } \leq \bigg[ \frac{ (1-e^{-\beta})}{(1-e^{-\beta'})} e^{(\beta' - \beta) \Delta_{AH}} \bigg] \bigg[(1-e^{-\beta'}) \sum_{\Delta_{A} \geq  \Delta_{AH}} e^{-\beta' \Delta_{A} } \bigg]~. \label{bound2}
\end{equation}
In order to find a common factor between the two bounds \eqref{bound1}-\eqref{bound2}, we restrict the thresholds $\Delta_{AH}$ and $\Delta_{BH}$ such that
\begin{equation}\label{preineq}
    \frac{ (1-e^{-\beta})}{(1-e^{-\beta'})} e^{(\beta' - \beta) \Delta_{AH}}\leq e^{(\beta' - \beta) \Delta_{BH}}~.
\end{equation}
Using the monotonicity of the exponential function, we see that the above inequality holds provided that
\begin{equation} \label{DeltaAH-DeltabHoffset}
    \Delta_{AH} - \Delta_{BH} \geq \frac{1}{(\beta - \frac{4 \pi^{2}}{\beta})}\log\left(\frac{1-e^{-\beta}}{1-e^{-4 \pi^{2}/\beta}}\right)\equiv g(\beta)~.
\end{equation}
To apply \eqref{preineq} for all $\beta>2\pi$ we must therefore choose $\Delta_{AH}-\Delta_{BH}$ to be larger than the maximum value of the function $g(\beta)$ in the range $2\pi <\beta <\infty.$ A plot of $g(\beta)$ is shown in Figure \ref{gbeta}.  Henceforth we thus choose the threshold $\Delta_{AH}$ in the definition \eqref{zheavy} of $Z_{H}$ as:
\begin{equation}\label{gstardef}
    \Delta_{AH}=\Delta_{BH}+g_{*}~,
\end{equation}
where $g_{*} \approx 0.0123...$ is the maximum value of $g(\beta)$.  Then, using \eqref{bound1}-\eqref{preineq} we see that we can bound $Z_{H}(\beta)$ as:
\begin{equation}
    Z_{H}(\beta) \leq \frac{e^{\beta\frac{(c - 1)}{12}}}{|\eta(\frac{i\beta}{2\pi})|^{2}} e^{(\beta' - \beta) \Delta_{BH}}\bigg[ (1-e^{-\beta'})\sum_{\Delta_{A} \geq  \Delta_{BH}+g_{*}} e^{-\beta' \Delta_{A} } + \sum_{\Delta_{B} \geq \Delta_{BH}} e^{-\beta' \Delta_{B}} \bigg]~.
\end{equation}
Using further the modular property of the Dedekind $\eta$-function $\eta(-1/\tau) = \sqrt{-i\tau} \eta(\tau)$, we can write this as an inequality relating the heavy partition function and its modular transform
\begin{equation} \label{hardboundedbymodularhard}
    Z_{H}(\beta) \leq \frac{\beta}{2\pi} e^{-(\beta - \beta')(\Delta_{BH} - \frac{(c - 1)}{12})} Z_{H}(\beta')~, \hspace{.2in} \beta>2\pi~.
\end{equation}

Having arrived at the inequality \eqref{hardboundedbymodularhard}, we can now proceed directly as in \cite{Hartman:2014oaa}.  First, we subtract $Z_{H}(\beta')$ on both sides of \eqref{hardboundedbymodularhard}:
\begin{equation}\label{stepa}
    Z_{H}(\beta) - Z_{H}(\beta') \leq -\Big(1-\frac{\beta}{2\pi}e^{-(\beta - \beta')(\Delta_{BH} - \frac{(c - 1)}{12})}\Big)Z_{H}(\beta')~.
\end{equation}
Next, we recall that modular invariance implies:
\begin{equation}\label{stepb}
    Z_{L}(\beta) + Z_{H}(\beta) = Z_{L}(\beta') + Z_{H}(\beta')~.
\end{equation}
Therefore, provided that 
\begin{equation}\label{dbhrest}
    \Big(1-\frac{\beta}{2\pi}e^{-(\beta - \beta')(\Delta_{BH} - \frac{(c - 1)}{12})}\Big)>0~,
\end{equation}
we can combine \eqref{stepa} and \eqref{stepb} to show that:
\begin{equation}
    Z_{H}(\beta') \leq \frac{Z_{L}(\beta) - Z_{L}(\beta')}{\Big(1-\frac{\beta}{2\pi}e^{-(\beta - \beta')(\Delta_{BH} - \frac{(c - 1)}{12})}\Big)}\leq  \frac{Z_{L}(\beta)}{\Big(1-\frac{\beta}{2\pi}e^{-(\beta - \beta')(\Delta_{BH} - \frac{(c - 1)}{12})}\Big)}~.
\end{equation}
Now using \eqref{hardboundedbymodularhard} we obtain:
\begin{equation}
    Z_{H}(\beta) \leq \frac{\frac{\beta}{2\pi} e^{-(\beta - \beta')(\Delta_{BH} - \frac{(c - 1)}{12})}}{\Big(1-\frac{\beta}{2\pi}e^{-(\beta - \beta')(\Delta_{BH} - \frac{(c - 1)}{12})}\Big)}  Z_{L}(\beta)~,
\end{equation}
and adding $Z_{L}(\beta)$ on both sides we obtain the desired bound on the torus partition function:
\begin{equation} \label{HKS-likebound}
    Z_{T^{2}}(\beta) \leq \frac{Z_{L}(\beta)}{1 - \frac{\beta}{2\pi} e^{-(\beta - \beta')(\Delta_{BH} - \frac{(c - 1)}{12})}}~,
\end{equation}
where explicitly, the light-state partition function $Z_{L}(\beta)$ is:
\begin{equation} \label{light-partition-function}
    Z_{L}(\beta) = \frac{e^{\beta\frac{(c - 1)}{12}}}{|\eta(\frac{i\beta}{2\pi})|^{2}} \Big[ (1-e^{-\beta})^{2} + (1-e^{-\beta})\sum_{\Delta_{A} < \Delta_{BH} + g_{*}} e^{-\beta \Delta_{A}} + \sum_{\Delta_{B} < \Delta_{BH}} e^{-\beta \Delta_{B}}  \Big]~.
\end{equation}

To determine the allowed heavy state threshold $\Delta_{BH}$ we investigate the assumption \eqref{dbhrest}.  Clearly, this is satisfied provided that for all $\beta>2\pi$
\begin{equation}
    \Delta_{BH} >  \frac{\log(\beta/2\pi)}{\beta-4\pi^{2}/\beta}+\frac{(c - 1)}{12} ~.
\end{equation}
As a function of $\beta,$ the right-hand side is maximized as $\beta\rightarrow 2\pi$ and evaluating at that point we find that it is sufficient to choose the threshold $\Delta_{BH}$ to obey
\begin{equation} \label{delta}
   \Delta_{BH} > \frac{(c - 1)}{12}+ \frac{1}{4\pi}~.
\end{equation}

The inequality \eqref{HKS-likebound} is the most general result of this subsection.  However, it is conceptually convenient to also state a coarser inequality that is less sensitive to the dimensions of light primary operators.  To do so, we note that the contribution of each primary to the quantity in square brackets appearing in \eqref{light-partition-function} is weighted by a number which is strictly less than one.  Therefore we can also write:
\begin{equation} \label{HKS}
    Z_{T^{2}}(\beta) \leq \frac{e^{\beta\frac{(c - 1)}{12}}}{(1 - \frac{\beta}{2\pi} e^{-(\beta - \beta')(\Delta_{BH} - \frac{(c - 1)}{12})})|\eta(\frac{i\beta}{2\pi})|^{2}} N_{\Delta_{BH}}~,
\end{equation}
where $N_{\Delta_{BH}}$ is an integer which counts the number of light Virasoro primaries i.e.\ those appearing in the sums in \eqref{light-partition-function}.

\subsection{Bounds on Interval Partition Functions}\label{secintbound}

In this subsection we derive an inequality on the interval partition function of a rational CFT with boundary conditions that preserve (half of) the extended chiral algebra in terms of the bulk torus partition function.  (See Appendix \ref{appendixirrationaltheories} for some related discussion of irrational theories).

As in the review of section \ref{bcftreviewsec}, we consider a rational CFT on an interval with boundary conditions $A$ and $B$ obeying the extended chiral algebra gluing conditions \eqref{extendedgluing}, so that the partition function is a finite sum over chiral algebra characters:
\begin{equation} \label{ABpartition}
    Z_{AB}(\beta) = \sum_{i} n_{AB}^{i} \chi_{i}(\beta)~.
\end{equation}
To obtain an inequality on the above, we use modular invariance \eqref{closedchannelpartition} to write $Z_{AB}$ as a transition amplitude of boundary states. Introducing a resolution of the identity in terms of the orthonormal basis appearing in the Ishibashi states \eqref{Ishibashi-States}:
\begin{equation}
    Z_{AB}(\beta)= \{ \Theta B | \tilde{q}^{\frac{1}{2}(L_{0} + \overline{L}_{0}-c/12)} | A \} = \hspace{-.1in}\sum_{\substack{i,i', N_{i}, N'_{i}}} \{ \Theta B | i,N_{i} \rangle \otimes \overline{|i', N'_{i} \rangle} \langle i,N_{i} | \otimes \overline{\langle i', N'_{i} |} \tilde{q}^{\frac{1}{2}(L_{0} + \overline{L}_{0}-c/12)} |A \}~, \label{Introducing-Resolutions-of-Identity}
\end{equation}
where as usual, $\tilde{q} = e^{- 4 \pi^{2} / \beta}$.  It is easy to see that the various inner products localize the sums.  When the dust settles, we obtain an expression for the partition function expressed in terms of the coefficients $\mathcal{B}_{A}^{\ i}$ defined in \eqref{bdystates} which characterize the expansion of the boundary state in terms of Ishibashi states:
\begin{equation} \label{closed-overlap}
    Z_{AB}(\beta)= \sum_{i} \mathcal{B}_{B}^{\ i^{+}} \mathcal{B}_{A}^{\ i} \,e^{- \frac{2 \pi^{2}}{\beta}(2h_{i}-\frac{c}{12})} \sum_{N=0}^{\infty} D_{i}(N) e^{- \frac{2 \pi^{2}}{\beta} (2N)}~,
\end{equation}
where now the sum over $i$ runs over (a subset of all) Virasoro scalar states (recall that the Ishibashi states \eqref{Ishibashi-States} are expanded in terms of Virasoro scalars), and $D_{i}(N)$ keeps track of any degeneracies of modes at level $N$ in the representation labeled by $i$.

We can now obtain an inequality on the interval partition function by bounding the expansion coefficients $\mathcal{B}_{A}^{\ i}$ in terms of quantities which are independent of the index $i$, but may depend on the external index $A$ labeling the boundary condition.  Specifically, we use the Cardy condition \eqref{Cardycondition} together with the unitarity of the $S$ matrix to obtain
\begin{equation} \label{inequalityfromCardyCond}
    \big| \mathcal{B}^{\ i^{+}}_{B} \mathcal{B}^{\ i}_{A} \big| = \big| \sum_{j} n^{j}_{AB} S^{-1}_{ji} \big| \leq   \sum_{j} |n^{j}_{AB}|< \infty~.
\end{equation}
Note that we have used the rationality, in particular the finite range of the primary index $j$, to assert that the final sum is finite.  Using this, the interval partition function \eqref{closed-overlap} is bounded as:
\begin{equation} \label{first-bound}
     Z_{AB}(\beta) \leq  \Big( \sum_{j} |n^{j}_{AB}| \Big) \bigg( \sum_{i} e^{- \frac{2 \pi^{2}}{\beta}(2h_{i}-\frac{c}{12})} \sum_{N=0}^{\infty} D_{i}(N) e^{- \frac{2 \pi^{2}}{\beta} (2N)} \bigg)~.
\end{equation}
The previous result is valid for both diagonal and non-diagonal theories.\footnote{We can also consider an overlap $\{ \Theta B | \tilde{q}^{\frac{1}{2}(L_{0} + \overline{L}_{0}-c/12)} | A_{\Omega} \}$ where one of the boundary states preserves a symmetry with a non-trivial automorphism $\Omega$. Then, in computing the overlap as in \eqref{Introducing-Resolutions-of-Identity} only representations that are invariant under $\Omega$ contribute in the sum. Thus, we still obtain the bound \eqref{first-bound}.} If we assume we are working in a diagonal RCFT we can also use the explicit Cardy solution \eqref{cardystates}:
\begin{equation} \label{diagonalRCFTcoefficientbound1}
  \mathcal{B}^{\ i}_{a} = \frac{S_{ai}}{\sqrt{S_{0i}}} = S_{ai} \sqrt{\frac{S_{00}}{S_{0i}}} \frac{1}{\sqrt{S_{00}}} = S_{ai} \sqrt{\frac{\mathcal{D}}{d_{i}}}~,
\end{equation}
where we have introduced the quantum dimension of each primary $d_{i}$ and the total quantum dimension $\mathcal{D}$:
\begin{equation}
    d_{i}=\frac{S_{0i}}{S_{00}}~, \hspace{.2in}\mathcal{D}=\frac{1}{S_{00}}~.
\end{equation}
Using the fact that for any representation, the quantum dimension $d_{i}\geq 1,$ (see e.g. \cite{Fuchs:1993et} and references therein) equation \eqref{diagonalRCFTcoefficientbound1} allows us to deduce that
\begin{equation}
    |\mathcal{B}^{\ i}_{a}|\leq \sqrt{\mathcal{D}}~.
\end{equation}
Therefore, for rational diagonal CFTs, the bound \eqref{first-bound} instead reads
\begin{equation} \label{first-bounddiag}
     Z_{ab}(\beta) \leq  \mathcal{D} \bigg( \sum_{i} e^{- \frac{2 \pi^{2}}{\beta}(2h_{i}-\frac{c}{12})} \sum_{N=0}^{\infty} D_{i}(N) e^{- \frac{2 \pi^{2}}{\beta} (2N)} \bigg)~.
\end{equation}

We can now compare the right-hand side of \eqref{first-bound} or \eqref{first-bounddiag} to the bulk torus partition function.  Up to the factor $\sum_{j} n_{AB}^{j},$ or $\mathcal{D}$ we recognize the expression as the torus partition function $Z_{T^{2}}(2\pi^{2}/\beta)$ truncated to a sum over (a subset of) scalar Virasoro states only. Adding back the contributions from the other operators we therefore obtain in general
\begin{equation} \label{bcft-bound}
    Z_{AB}(\beta) \leq  \Big( \sum_{j} |n^{j}_{AB}| \Big) Z_{T^{2}}\Big( \frac{2\pi^{2}}{\beta}\Big) = \Big( \sum_{j} |n^{j}_{AB}| \Big) Z_{T^{2}} ( 2 \beta )~,
\end{equation}
or, in the case of rational diagonal theories, 
\begin{equation} \label{bcft-bounddiag}
    Z_{ab}(\beta) \leq  \mathcal{D} Z_{T^{2}}\Big( \frac{2\pi^{2}}{\beta}\Big) = \mathcal{D} Z_{T^{2}} ( 2 \beta )~.
\end{equation}

\subsection{Inequalities Relating Boundary and Bulk Data}\label{groundstatedeg}

Our task is now to combine the interval bounds \eqref{bcft-bound}-\eqref{bcft-bounddiag} derived above, with the bulk inequalities stated in \eqref{HKS-likebound} and \eqref{HKS}.  For general rational theories, we have for $\beta>\pi$ 
\begin{equation}\label{inthksnondiag}
     Z_{AB}(\beta) \leq \Big( \sum_{j} |n^{j}_{AB}| \Big) \frac{e^{\beta \frac{(c - 1)}{6}}}{\Big(1 - \frac{\beta}{\pi} e^{-(2\beta - \frac{2\pi^{2}}{\beta})\big(\Delta_{BH} - \frac{(c - 1)}{12}\big)}\Big) \big| \eta(\frac{i\beta}{\pi}) \big|^{2}} N_{\Delta_{BH}}~,
\end{equation}
while for diagonal theories we instead have\footnote{Note that the cylinder partition function \eqref{first-bound} includes only scalar states and so one could have actually used an inequality on the bulk partition function truncated to consider these states only. The advantage of using \eqref{bcft-bound} is that in both \eqref{inthksnondiag} and \eqref{inthksdiag} one finds simple universal coefficients. }
\begin{equation}\label{inthksdiag}
 Z_{AB}(\beta) \leq \mathcal{D} \frac{e^{\beta \frac{(c - 1)}{6}}}{\Big(1 - \frac{\beta}{\pi} e^{-(2\beta - \frac{2\pi^{2}}{\beta})\big(\Delta_{BH} - \frac{(c - 1)}{12}\big)}\Big) \big| \eta(\frac{i\beta}{\pi}) \big|^{2}} N_{\Delta_{BH}}~.
\end{equation}
The results above may be used to constrain the degeneracies of each energy level in the Hilbert space $\mathcal{H}_{AB}.$  Indeed, each chiral algebra character may be expanded as 
\begin{equation}
    \chi_{i}(\beta) = e^{- \beta(h_{i} - c/24)}\sum_{n \geq 0} d_{i}(n) e^{- n\beta}~,
\end{equation}
where $h_{i}$ is the conformal weight of the primary associated with the representation labeled by $i$ and $d_{i}(n)$ counts possible degeneracies at level $n$. Since the interval partition function is a sum over such characters weighted by $n_{AB}^{i}$ (which are positive) we deduce that, for $\beta > \pi$
\begin{equation}
    n_{AB}^{i} d_{i}(0) q^{h_{i} - c/24} \leq Z_{AB}(\beta) \leq \Big( \sum_{j} |n^{j}_{AB}| \Big) \frac{e^{\beta \frac{(c - 1)}{6}}}{\Big(1 - \frac{\beta}{\pi} e^{-(2\beta - \frac{2\pi^{2}}{\beta})\big(\Delta_{BH} - \frac{(c - 1)}{12}\big)}\Big) \big| \eta(\frac{i\beta}{\pi}) \big|^{2}} N_{\Delta_{BH}}~,
\end{equation}
rearranging to isolate the degeneracy at a fixed level we therefore have for each $i$ and for all $\beta>\pi$:
\begin{equation} \label{bound}
    n_{AB}^{i} d_{i}(0) \leq \Big( \sum_{j} |n^{j}_{AB}| \Big) \frac{e^{\beta \big( \frac{c + 8 h_{i}}{8}-\frac{1}{6}\big)}}{\Big(1 - \frac{\beta}{\pi} e^{-(2\beta - \frac{2\pi^{2}}{\beta})\big(\Delta_{BH} - \frac{(c - 1)}{12}\big)}\Big) \big| \eta(\frac{i\beta}{\pi}) \big|^{2}} N_{\Delta_{BH}}~. 
\end{equation}
In particular, we may apply this to constrain the ground state degeneracy $d_{AB}$ discussed around equation \eqref{intgroundstate} with associated conformal weight $h_{AB,\text{min}}$:  
\begin{equation}\label{bndiag}
    d_{AB}\leq \Big( \sum_{j} |n^{j}_{AB}| \Big) \frac{e^{\beta \big( \frac{c + 8 h_{AB,\text{min}}}{8}-\frac{1}{6}\big)}}{\Big(1 - \frac{\beta}{\pi} e^{-(2\beta - \frac{2\pi^{2}}{\beta})\big(\Delta_{BH} - \frac{(c - 1)}{12}\big)}\Big) \big| \eta(\frac{i\beta}{\pi}) \big|^{2}} N_{\Delta_{BH}}~.
\end{equation}
Analogously, for a rational diagonal theory with Cardy boundary conditions the ground state degeneracy is bounded by 
\begin{equation}\label{bdiag}
    d_{ab}\leq \mathcal{D} \frac{e^{\beta \big( \frac{c + 8 h_{ab,\text{min}}}{8}-\frac{1}{6}\big)}}{\Big(1 - \frac{\beta}{\pi} e^{-(2\beta - \frac{2\pi^{2}}{\beta})\big(\Delta_{BH} - \frac{(c - 1)}{12}\big)}\Big) \big| \eta(\frac{i\beta}{\pi}) \big|^{2}} N_{\Delta_{BH}}~.
\end{equation}

The bounds above are valid for any $\beta>\pi$ and for any threshold $\Delta_{BH}$ obeying the constraint \eqref{delta}.  In practical terms, one is often ignorant about the degeneracy of light operators.  Therefore to apply the inequalities, it is convenient to fix a particular value of $\Delta_{BH}$.  Below, we therefore choose:
\begin{equation}\label{simpthresh}
    \Delta_{BH}\rightarrow \frac{c-1}{12}+\frac{1}{2\pi}~,
\end{equation}
and correspondingly define $N$ to be the light primary count $N_{\Delta_{BH}}$ defined below \eqref{HKS} at the above value of $\Delta_{BH}.$  Introducing
\begin{equation}
    c_{\mathrm{eff}} \equiv c+ 8h_{ab,\text{min}}~,
\end{equation}
the diagonal bound \eqref{bdiag} then reads for instance:
\begin{equation}\label{bdiag1}
    \frac{d_{ab}}{\mathcal{D}N}\leq  \frac{e^{\beta \big( \frac{c_{\mathrm{eff}}}{8}-\frac{1}{6}\big)}}{\Big(1 - \frac{\beta}{\pi} e^{-(\frac{\beta}{\pi} - \frac{\pi}{\beta})}\Big) \big| \eta(\frac{i\beta}{\pi}) \big|^{2}} ~.
\end{equation}
Since \eqref{bdiag1} is valid for any $\beta > \pi$ we can now optimize this inequality by minimizing the right-hand side with respect to $\beta$.  Denoting the resulting minimum value by $f(c_{\mathrm{eff}})$ we have therefore derived a bound:
\begin{equation}\label{finbndd}
    \frac{d_{ab}}{\mathcal{D}N}\leq f(c_{\mathrm{eff}})~,
\end{equation}
where the function $f(c_{\mathrm{eff}})$ is plotted in Figure \ref{figiplot}.  Analogously, for rational non-diagonal theories, we have the bound:
\begin{equation}\label{finbnddn}
    \frac{d_{AB}}{\left(\sum_{j} n^{j}_{AB}\right)N}\leq f(c_{\mathrm{eff}})~.
\end{equation}

Equations \eqref{finbndd} and \eqref{finbnddn} are the main results of this section.  They provide a priori constraints on the ground state degeneracy of any rational CFT on the interval in terms of closed sector data.  As some further comments we observe:
\begin{itemize}
    \item For large central charge, $c_{\text{eff}}\rightarrow \infty$, the optimization over $\beta$ may be done explicitly in a $1/c_{\text{eff}}$ expansion.  In this case it is straightforward to see that the minimum of the function appearing on the right-hand side of \eqref{bdiag1}, occurs near the boundary $\beta\rightarrow \pi,$ hence we expand
    \begin{equation}
        \beta=\pi +\frac{x}{c_{\text{eff}}}+\cdots~,
    \end{equation}
    where the neglected terms are higher order in $1/c_{\text{eff}}$.  Optimization now leads to an order one value for the coefficient $x$ and hence the bound \eqref{finbndd} reads:
    \begin{equation} \label{largecbound}
       \log\left(\frac{d_{ab}}{\mathcal{D}N}\right)\leq \frac{\pi}{8}c_{\text{eff}}+\log(c_{\text{eff}})+\text{finite}~,
    \end{equation}
    with a similar statement for the non-diagonal case,
      \begin{equation}\label{asymp}
       \log\left(\frac{d_{AB}}{\left(\sum_{j} n^{j}_{AB}\right)N}\right)\leq \frac{\pi}{8}c_{\text{eff}}+\log(c_{\text{eff}})+\text{finite}~.
    \end{equation}
    Bounds of this qualitative form,  however without the additional factors of the total quantum dimension $\mathcal{D}$ and light operator degeneracy $N$, was considered in \cite{2017}, and conjectured to hold even at small $c$. (See footnote \ref{footcounter}.) By contrast our result derived rigorously from modular invariance only asymptotically takes this form for large central charge and suppresses the ground state degeneracy by the factor $\mathcal{D}N$ (which may also grow with $c_{\text{eff}}.$)  
    \item For small central charge, the operators counted by $N_{\Delta_{BH}}$ appearing in the bounds above are very light.  For instance taking $\Delta_{BH}$ as in \eqref{simpthresh}, the Virasoro primaries counted by $N$ appearing in \eqref{finbndd}-\eqref{finbnddn} are all strictly relevant operators provided that:
    \begin{equation}
        c< 25-\frac{6}{\pi}-12g_{*}\approx 22.94~.
    \end{equation}
\end{itemize}

\subsection{Example: Inequalities for $SU(2)_{k}$ at Large $k$}

Let us illustrate the features of the inequalities derived above in the class of examples $SU(2)_{k}$ in the limit of large level $k$. In this theory, the central charge and total quantum dimension are:
\begin{equation}
    c=\frac{3k}{k+2}\approx 3~, \hspace{.2in}\mathcal{D}=\frac{1}{S_{00}}=\sqrt{\frac{k+2}{2}}\sin\left(\frac{\pi}{k+2}\right)^{-1}\approx \frac{k^{3/2}}{\sqrt{2}\pi}~.
\end{equation}
As is well known (see e.g.\ \cite{DiFrancesco:639405}), this model has holomorphic Kac-Moody primaries labelled by a half integral spin $j$ with range $0\leq j \leq k/2$.  The associated conformal weights are
\begin{equation}
    h_{j}=\frac{j(j+1)}{k+2}~.
\end{equation}
Note that in this case the threshold $\Delta_{BH}$ specified by \eqref{simpthresh} is less than one, and therefore we do not need to consider any Kac-Moody descendants other than those related by the action of zero modes.  The maximum spin $j_{\text{max}}$ contributing to the light operators is then easily determined, and the associated Virasoro primary count $N$ is given by summing over products of holomorphic and antiholomorphic primaries in representations of $SU(2)$ up to spin $j_{\text{max}}$.  Hence
\begin{equation}
     j_{\text{max}} \sim\sqrt{k}~, \hspace{.2in}N\sim \sum_{\ell=0}^{j_{\text{max}}}\ell^{2} \sim k^{3/2}~,
\end{equation}
where above we have neglected positive order one coefficients.

We now consider the Cardy boundary conditions $a=0$ (i.e.\ associated to the identity), and $b=k^{\alpha}$ associated to an arbitrary primary whose spin grows as fractional power of $k$ with $1/2 <\alpha\leq 1$).  The fusion coefficients then satisfy $N_{0b}^{i}=\delta_{b}^{i}$ so we find a single character in the interval partition function $Z_{0b}(\beta)$.  The ground state degeneracy and effective central charge are then:
\begin{equation}
    d_{0b}\sim k^{\alpha}~, \hspace{.2in}c_{\text{eff}}\sim k^{2\alpha-1} \gg 1~.
\end{equation}
Neglecting order one numbers, and using the fact that the sum of fusions coefficients $\sum_{j}N_{0b}^{j}$ is one, the bound \eqref{asymp} reads
\begin{equation}
\log\left(\frac{d_{AB}}{\left(\sum_{j} n^{j}_{AB}\right)N}\right)\sim\log\left(\frac{k^{\alpha}}{k^{3/2}}\right)\leq c_{\text{eff}}\sim k^{2\alpha-1}~,
\end{equation}
which is clearly satisfied for any $\alpha>1/2.$ 

\section{Boundary Conditions for $SU(M)_{1}$ WZW and Bulk Deformations} \label{section4}

In this section we explore the bounds derived in section \ref{section3} and the map between bulk relevant deformations and boundary states discussed in section \ref{subsection2.2}, in the context of $SU(M)_{1}$ WZW models. 

\subsection{Bound at Small and Large $M$} \label{subsubsection4.2.1}

Let us begin by demonstrating the central charge inequalities \eqref{finbndd}-\eqref{finbnddn} for $SU(M)_{1}$ theory for small $M$. The central charge and total quantum dimension are:
\begin{equation} \label{SUM1centrlachargeandD}
    c=M-1~, \hspace{.3in}\mathcal{D}=\frac{1}{S_{00}}=\sqrt{M}~,
\end{equation}
and the holomorphic conformal weights of the chiral algebra primaries are given by
\begin{equation} \label{sum1conformalweights}
     h_{i} = \frac{i(M-i)}{2M}~, \quad i = 1, \ldots ,M - 1~.
\end{equation}
The associated scaling dimensions are then $\Delta_{i} = h_{i} + \overline{h}_{i} = \frac{i(M-i)}{M}$. To apply our inequalities we must determine the number $N$ of Virasoro primaries below the threshold \eqref{simpthresh} which for these theories is:
\begin{equation}
    \Delta_{BH}\rightarrow \frac{M-2}{12}+\frac{1}{2\pi}~.
\end{equation}
We then observe that for $M$ small enough ($M \leq 10$), all Kac-Moody descendants and and all non-identity Kac-Moody primaries have scaling dimension above this threshold.  Therefore, for these small values of $M,$ the identity is the only light operator and hence $N=1$.

Next we turn to the boundary states and ground state degeneracies.  As an illustrative example, we choose the Cardy boundary conditions corresponding to the identity ($a=0$) and a non-trivial primary ($b=i$). As the only non-trivial fusion coefficient is $N_{0i}^{i}=1$ the associated cylinder partition function \eqref{Cardy-OpenPartition} is simply:
\begin{equation}
    Z_{0i}(q)=\chi_{i}(q)~.
\end{equation}
In particular, the ground state degeneracy for these boundary conditions is the degeneracy of the $i$-th chiral primary
\begin{equation}
    d_{0i} = {M \choose i}~.
\end{equation}
Finally we can also read off $c_{\mathrm{eff}}$ from \eqref{SUM1centrlachargeandD} and \eqref{sum1conformalweights}:
\begin{equation}
   c_{\mathrm{eff}} =M-1+ \frac{4i(M-i)}{M}~.
\end{equation}
We can now verify the bounds \eqref{finbndd}-\eqref{finbnddn}. For small $M$, this leads to the data points shown in Figure \ref{figiplot}, which comfortably lie in the allowed region.

We turn now to the analysis of the bound in the regime of large $M$. For simplicity we take $M=2m$ even, and study the transition where the boundary conditions are labelled by the identity and the primary corresponding to the middle fundamental weight $\mathbf{w}_{m}$, which is the choice that maximizes the ground state degeneracy. As we are interested in large $M,$ we use the large central charge bound \eqref{largecbound} and consider only the leading contributions. 

With the previous setup, the degeneracy of the ground state is
\begin{equation}
    d_{0m} = {{2m}\choose{m}} \Longrightarrow \ln{(d_{0m})} \approx M \ln{2}~.
\end{equation}
From \eqref{SUM1centrlachargeandD} and \eqref{sum1conformalweights} it is also easy to see that $c_{\mathrm{eff}} \approx 2M$ at large $M$. Then, from \eqref{largecbound} we have:
\begin{equation}\label{asymex}
   \log\left(\frac{d_{ab}}{\mathcal{D}N}\right)\leq \frac{\pi}{8}c_{\text{eff}} \longleftrightarrow  M \ln{2} < \frac{\pi}{4}M + \log(N)~,
\end{equation}
up to terms that grow as $\log(M)$. The inequality \eqref{asymex} is satisfied irrespective of the asymptotic growth of the light primary degeneracy $N$.\footnote{Note that the non-zero module Virasoro primaries of $SU(M)_{1}$ of scaling dimension less than $\Delta_{BH}$ correspond to the integer points interior to an ellipse defined by the $(2M-2)$ winding and momentum quantum numbers of characteristic radius $\sqrt{\Delta_{BH}}$, given by the heavy threshold $\Delta_{BH}\sim M$. An estimation on the number of such lattice points has been studied \cite{Mazo1990LatticePI} in the related case of an $M$-dimensional sphere with characteristic radius $\sim \sqrt{M}$ where the logarithm of the number of lattice points inside the sphere grows linearly in $M$ at large $M$.
Analogously, we expect the growth the ellipse-case to be such that $N\sim \log(M).$}

\subsection{Bulk Deformations and Boundary Conditions}\label{su2case-subsection} 

In this section we directly deform the $SU(M)_{1}$ theory by relevant operators to flow to trivially gapped phases.  To place this in the context of symmetry protected renormalization group flows described in section \ref{sec:symflow}, we recall that the global symmetry of the WZW model $SU(M)_{1}$ is 
\begin{equation}
    \frac{(SU(M)\times SU(M))}{\mathbb{Z}_{M}} \rtimes \mathbb{Z}_{2}~,
\end{equation}
where the quotient is by the common center, and the semidirect product is charge conjugation.\footnote{For the special case of $M=2$ there is no charge conjugation symmetry.} We will investigate deformations that preserve a $SU(M)/\mathbb{Z}_{M}$ subgroup of the symmetry at the critical point and flow to various symmetry protected phases.  

Below, we will study these models by making use of the well-known description of the $SU(M)_{1}$ WZW theory in terms of $M-1$ free bosons.  In general, for a simply-laced algebra of rank $\ell$ this description involves $\ell$ bosons $\mathbf{X}$ compactified on the corresponding root lattice. (Recall for any such lattice there are $\ell$ simple roots $\mathbf{r}_{j}$ and $\ell$ fundamental weights $\mathbf{w}_{i}$ such that $\mathbf{w}_{i} \cdot \mathbf{r}_{j} = \delta_{ij}$.) For the $A_{M-1}$ algebra case, the fundamental weights are associated to vertex operators $e^{i \mathbf{w}_{i} \cdot \mathbf{X}}$ with scaling dimensions
\begin{equation} \label{sunprimaries}
     \Delta_{\mathbf{w}_{i}} = \frac{i(M-i)}{M}~, \quad i = 1 \ldots M - 1~,
\end{equation}
where $\Delta_{\mathbf{w}_{i}} < 2$ for relevant deformations. In the following we will concentrate on the most-relevant deformations corresponding to $\Delta_{\mathbf{w}_{1}} = \Delta_{\mathbf{w}_{M-1}} = (M-1)/M$. As we will see, these will be enough to drive flows to various trivially gapped phases. Related work on constructing SPT transitions protected by $PSU(M)$ symmetry from Hamiltonian models can be found in \cite{PhysRevLett.59.799, PhysRevB.93.165135, Nonne_2013, Roy:2015ars}.

Since the flows below result in trivially gapped phases they are described by SPTs.  As the preserved global symmetry is $PSU(M),$ the relevant bosonic SPTs are labelled by:
\begin{equation}
    H^{2}(SU(M)/\mathbb{Z}_{M})\cong \mathbb{Z}_{M}~.
\end{equation}
Abstractly, we may describe these using the second Stiefel-Whitney class $w_{2}(A) \in H^{2}(X,\mathbb{Z}_{M})$.  Here $X$, is spacetime and $A$ indicates a background $PSU(M)$ connection.  This cohomology class measures the obstruction to lifting a $PSU(M)$ bundle to an $SU(M)$ bundle.  The $p$-th SPT is then defined by a partition function:
\begin{equation} \label{SPTw2}
     \exp{\left( \frac{2 \pi i p}{M} \int_{X} w_{2}(A) \right)}~,
\end{equation}
where $p\sim p+M$.  In the free boson description, only the symmetry associated to the Cartan subalgebra $U(1)^{M-1}$ is manifest.  However, by activating background gauge fields for the Cartan subalgebra, we can recover complete information about the SPT phase \eqref{SPTw2}.  Specifically, for the choice of Cartan subalgebra described in subsection \ref{subsection4.2.2} below, we have:
\begin{equation}\label{w2corr}
    \int w_{2}(A)\bigg|_{A \, \in \, \mathrm{Cartan}} \longrightarrow  \int\sum_{j} \frac{j dA^{j}}{2\pi}~.
\end{equation}
So, the SPT action \eqref{SPTw2} reduces to
\begin{equation}
 \exp{\left(\frac{2 \pi i p}{M} \int w_{2}(A) \right)}\Bigg|_{A \, \in \, \mathrm{Cartan}} \longrightarrow \exp{\Bigg( \frac{2 \pi i p}{M} \int \sum_{j} \frac{ j dA^{j}}{2\pi} \Bigg)}~,
\end{equation}
where the sum over $j$ runs over the rank of the $su(M)$ algebra, and the fluxes are quantized as $\frac{1}{2\pi} \int dA^{j} \in \mathbb{Z}$.

\subsubsection{Deforming the $SU(2)_{1}$ Critical Theory and SPT Phases}\label{su2case-subsection}

In this section we illustrate the appearance of different SPT phases in the simple case of $SO(3)\cong SU(2)/\mathbb{Z}_{2}$ symmetry, starting from the $SU(2)_{1}$ WZW critical theory and deforming it by the $\mathrm{Tr}(g)$ deformation:\footnote{In this case one can see the quotient by the common center by recalling that the Hilbert space for $SU(2)_{1}$ is $\mathcal{H} = \mathcal{H}_{0} \otimes \mathcal{H}_{0} + \mathcal{H}_{1/2} \otimes \mathcal{H}_{1/2}$  and hence transforms trivially under a $(-\id,-\id) \in SU(2) \times SU(2)$. }
\begin{equation} \label{su2wzw}
    S = \frac{1}{16\pi} \int d^{2}x \mathrm{Tr}(\partial^{\mu} g^{-1} \partial_{\mu}g)-i\frac{1}{24\pi}\int_{B} d^{3}x \, \epsilon_{\alpha \beta \gamma} \tilde{g}^{-1}\partial^{\alpha}\tilde{g} \, \tilde{g}^{-1}\partial^{\beta} \, \tilde{g}\tilde{g}^{-1}\partial^{\gamma}\tilde{g} + \lambda \int d^{2}x \, \mathrm{Tr}(g)~,
\end{equation}
where $g \in SU(2)$.  This deformation and the associated SPT transition (interpreted as an anomaly of the domain wall degrees of freedom) has been previously discussed in \cite{Chen_2013}.  In anticipation of the higher rank generalization, we will instead proceed using the dual compact boson description.

To show that this class of flows describes an SPT transition, we couple the theory to a background gauge field $A$ in the Cartan subalgebra $U(1)\subset SO(3)$. The free boson description consists of a single scalar at the self T-dual radius\footnote{Here we are using conventions where $\alpha' = 2$, so the self T-dual radius lies at $R= \sqrt{2}$.} $X \sim X + 2 \pi \sqrt{2}$ with an appropriate relevant deformation, and a background field $A$ coupled to the winding current $j^{w}_{\mu} = \epsilon_{\mu \nu} \partial^{\nu} X$. The full action takes the form:
\begin{equation} \label{gaugedaction}
    S[A,\lambda] = \frac{1}{8 \pi}\int d^{2}x \, \gamma^{ab}  \partial_{a}X\partial_{b}X -i \frac{1}{2 \pi \sqrt{2}}\int X \, dA    + \lambda \int d^{2}x \cos{(X/\sqrt{2})}~.
\end{equation}
The normalization in front of the second term in \eqref{gaugedaction} can be obtained by requiring single-valuedness of the path-integral under $X \rightarrow X + 2 \pi \sqrt{2}$, and using the quantization $\frac{1}{2\pi} \int dA \in \mathbb{Z}$.

The SPTs can be recovered in the extreme $\lambda \rightarrow \pm \infty$ limit. When $\lambda \ll 0$, $X \rightarrow 0$ is energetically favored and in the extreme $\lambda \rightarrow - \infty$ case the potential is infinitely deep and freezes $X=0$. Similarly, when $\lambda > 0$, $X \rightarrow \sqrt{2} \pi$ is favored, and in the extreme $\lambda \rightarrow \infty$ case the field is frozen at $X=\sqrt{2}\pi$. However, in the latter case we obtain a non-trivial contribution from the second term in \eqref{gaugedaction}. All in all, we observe we are in the presence of two different trivially gapped phases as described by the quotient of the partition functions:
\begin{equation} \label{SPTphaseSO3}
    \lim_{\lambda \rightarrow \infty} \frac{Z[A, +\lambda]}{Z[A, -\lambda]} = \exp \bigg({- i \pi \frac{1}{2 \pi}\int dA}\bigg)\rightarrow \exp\left(i\pi\int w_{2}(A) \right)~,
\end{equation}
where in the final step we have used the correspondence \eqref{w2corr} to restore the full $SO(3)$ SPT.  

Before moving on, it is useful to study  more carefully why the previous deformation works, which will pave the way for the analysis at higher rank. Recall that in the compactified boson at the self T-dual radius, the holomorphic currents generating (half of) the enhanced symmetry are given by
\begin{equation}
    j^{1}(z) = \cos{\big( \sqrt{2}X^{L}(z) \big)}~, \hspace{.2in}j^{2}(z) = \sin{\big( \sqrt{2}X^{L}(z) \big)}~, \hspace{.2in}j^{3}(z) = i \partial X^{L}(z)/\sqrt{2}~,
\end{equation}
with a corresponding expression for antiholomorphic currents in terms of $X^{R}(\overline{z})$. Since these currents only involve the holomorphic part, we have the following OPEs:
\begin{align}
e^{i\sqrt{2}X^{L}(z)} e^{-i\big(X^{L}(w) + X^{R}(\overline{w})\big)/\sqrt{2}} &\sim \frac{e^{i\big(X^{L}(w)-X^{R}(\overline{w})\big)/\sqrt{2}}}{(z-w)}~, \label{ope1} \\[0.4mm]
e^{-i\sqrt{2}X^{L}(z)} e^{i\big(X^{L}(w) + X^{R}(\overline{w})\big)/\sqrt{2}} &\sim \frac{e^{-i\big(X^{L}(w)-X^{R}(\overline{w})\big)/\sqrt{2}}}{(z-w)}~. \label{ope2}
\end{align}
Now, consider the combination of zero modes $J^{1}_{0} = j^{1}_{0}-\overline{j}^{1}_{0}$, $J^{2}_{0} = j^{2}_{0} + \overline{j}^{2}_{0}$, $J^{3} = j^{3}_{0} - \overline{j}^{3}_{0}$ generating an $A_{1}$ symmetry algebra.  Then from the previous OPEs, it is straightforward to show that the $\cos{(X/\sqrt{2})}$ deformation is annihilated by $(J^{1}_{0},J^{2}_{0},J^{3}_{0})$. This allows us to see purely from the compact boson description that the $\cos{(X/\sqrt{2})}$ deformation in \eqref{gaugedaction} indeed preserves a full $SO(3)$ symmetry along the flow, leading to the SPT phases obtained in \eqref{SPTphaseSO3}.

\subsubsection{Generalization to $SU(M)_{1}$} \label{subsection4.2.2}

For $A_{\ell}$ algebras of higher rank $\ell$ the free boson description involves $\ell$ fields $\mathbf{X}$ compactified on the corresponding root lattice. We now have $\ell$ winding currents and we introduce a background gauge field for each.  Following steps analogous to those of the $SU(2)_{1}$ case we obtain the action
\begin{equation} \label{gaugedactionhigherrank}
    S[A] = \frac{1}{8 \pi}\int d^{2}x \, \gamma^{ab}  \partial_{a}\mathbf{X} \cdot \partial_{b}\mathbf{X} -i \frac{1}{2 \pi}\int dA^{k} \, \mathbf{w}_{k} \cdot \mathbf{X}~.
\end{equation}
In the above,  the appearance of the weights $\mathbf{w}_{k}$ in the coupling to background fields can be understood from the fact that the weight lattice and root lattice are dual and thus these winding currents generate the minimally allowed charges.

Before setting-up concrete relevant deformations leading us to SPT transitions, it is instructive to see how the latter could arise from the action \eqref{gaugedactionhigherrank} in a similar spirit as they arose in the $SU(2)_{1}$ case. Clearly, the non-trivial phases must come from the dot products $\mathbf{w}_{k} \cdot \mathbf{X}$ in the gauging term. Now, notice that $\mathbf{w}_{k} \cdot \mathbf{w}_{j} =  A^{-1}_{(\ell) \, kj}$, with $A^{-1}_{(\ell)}$ the inverse of the Cartan matrix
\begin{equation} \label{inversecartanA}
    A^{-1}_{(\ell) \, kj} = \mathrm{min}(k,j) - \frac{kj}{(\ell+1)}~.
\end{equation}
Suppose that a relevant deformation were to localize $\mathbf{X} \rightarrow \mathbf{X}^{(j)} = 2 \pi \mathbf{w}_{j} $ for some $j$. This then would naturally realize an SPT in the deep IR.  Indeed, in that case the action \eqref{gaugedactionhigherrank} reduces to simply:
\begin{equation}\label{bigreduced}
 S[A]\longrightarrow   -i \int dA^{k} \, \mathbf{w}_{k} \cdot\mathbf{w}_{j} =-i A^{-1}_{(\ell) \, kj}\int dA^{k}=\frac{2\pi i j}{\ell+1}\int \sum_{k}\frac{k dA^{k}}{2\pi}\rightarrow \exp{\left(\frac{2 \pi i j}{\ell+1} \int w_{2}(A) \right)}~,
\end{equation}
where in the middle equations we have used flux quantization to drop the integral pieces of the inverse Cartan matrix, and in the last step we have used the correspondence \eqref{w2corr} to restore the full SPT. What is left then is to find a concrete physical realization; that is, a symmetry-preserving deformation leading to these trivially gapped phases.

We now exhibit explicit relevant deformations analogous to those in section \ref{su2case-subsection}. We are interested in the most-relevant allowed deformations, given by the representation associated to $\mathbf{w}_{1}$ and its conjugate. Unlike the rank one case, there are now two operators which preserve the same algebra. Specifically, both $C^{A_{\ell}}_{\mathbf{w}_{1}}$ and $S^{A_{\ell}}_{\mathbf{w}_{1}}$, with:
\begin{equation} \label{Vc}
    C^{A_{\ell}}_{\mathbf{w}_{1}} = \sum_{i=0}^{\ell} \cos{(\mathbf{w}_{i+1} \cdot \mathbf{X} - \mathbf{w}_{i} \cdot \mathbf{X})}~,
\end{equation}
\begin{equation} \label{Vs}
    S^{A_{\ell}}_{\mathbf{w}_{1}} = \sum_{i=0}^{\ell} \sin{(\mathbf{w}_{i+1} \cdot \mathbf{X} - \mathbf{w}_{i} \cdot \mathbf{X})}~,
\end{equation}
and $\mathbf{w}_{0} = \mathbf{w}_{\ell+1} = 0$, are annihilated by zero modes generating an $A_{\ell}$ algebra as in the rank one case. Thus, we can deform the action \eqref{gaugedactionhigherrank} by a general linear combination of \eqref{Vc} and \eqref{Vs} while preserving a fixed $PSU(M)$ symmetry. Notice that in the rank one case, since the representation $\mathbf{w}_{1}$ is self-conjugate one of the operators above is trivial. Indeed, if we apply the definition \eqref{Vs} to the rank one case we obtain $S^{A_{1}}_{\mathbf{w}_{1}} = \sin{(\mathbf{w}_{1} \cdot \mathbf{X})} + \sin{(-\mathbf{w}_{1} \cdot \mathbf{X})} = 0$.

In order to see that \eqref{Vc} and \eqref{Vs} preserve the same algebra one needs to be careful in how to appropriately define  both the currents and the operators $\eqref{Vc}$ and $\eqref{Vs}$ above. Indeed, the rank one discussion above in subsection \ref{su2case-subsection} was imprecise in that vertex operators of the compact boson theory at enhanced symmetry points must be ``dressed'' by correction factors in order for the vertex operators to appropriately generate the symmetry algebra and remove certain phases arising when interchanging vertex operators (see e.g.\ \cite{DiFrancesco:639405} section 15.6.3). For the purposes of streamlining the presentation the previous points are considered in Appendix \ref{appendixoncocycles} for the general rank case. As the analysis there illustrates, such correction factors do not play a substantial role in the discussion to follow.

\begin{figure}[t]
    \centering
    {{\includegraphics[width=8.04cm]{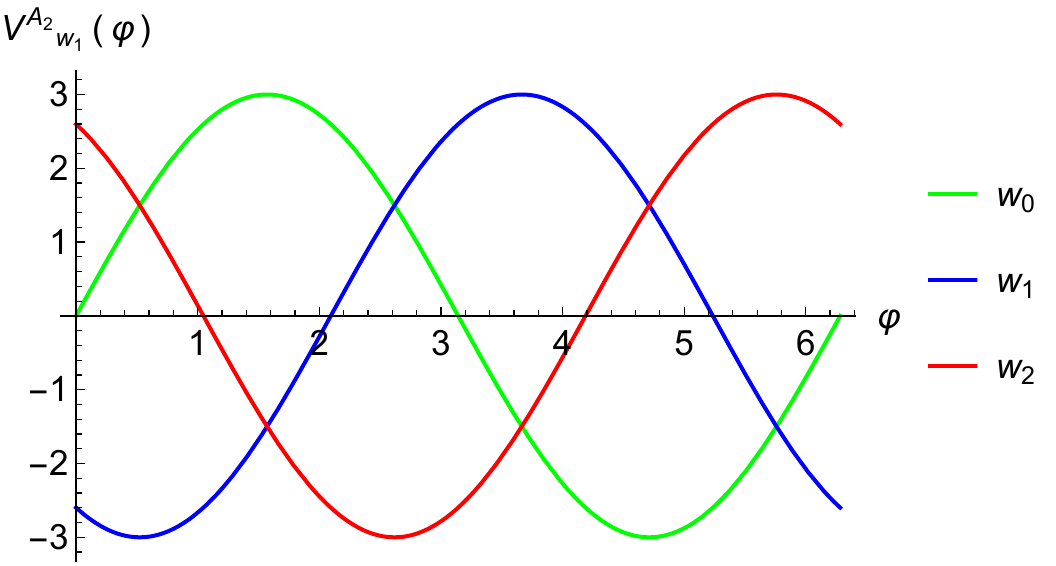} }}%
    \quad
    {{\includegraphics[width=8.04cm]{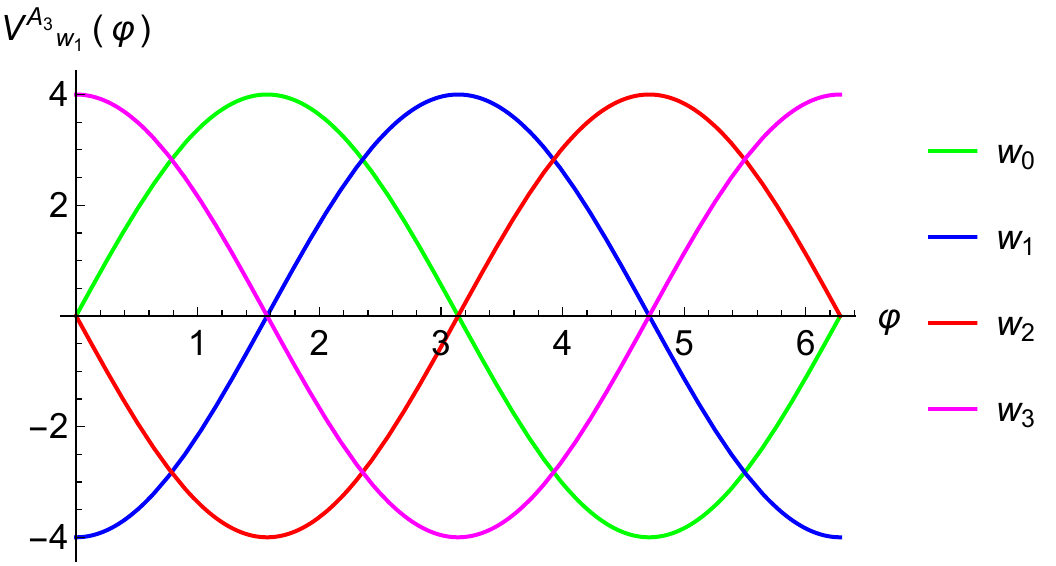} }}%
    \caption{The potential $V^{A_{\ell}}_{\mathbf{w}_{1}}(\varphi)$ at the critical points $\mathbf{X}^{(j)} = 2 \pi \mathbf{w}_{j}$ as a function of $\varphi$. Left: Illustration in the case $\ell=2$. Right: Illustration in the case $\ell=3$. For generic $\varphi$ the global maximum and minimum are unique.  At special $\varphi$ there is level crossing.}
    \label{fig:exampleplots}
\end{figure}

The most-relevant deformation thus results in a potential preserving a $PSU(M)$ symmetry, and it is of the form
\begin{align}
    V^{A_{\ell}}_{\mathbf{w}_{1}}(\varphi) &= \cos{(\varphi)}S^{A_{\ell}}_{\mathbf{w}_{1}} + \sin{(\varphi)}C^{A_{\ell}}_{\mathbf{w}_{1}} \nonumber \\ &= \sin{(\mathbf{w}_{1} \cdot \mathbf{X} + \varphi)} + \sum_{i=1}^{\ell-1} \sin{ (\mathbf{w}_{i+1} \cdot \mathbf{X} - \mathbf{w}_{i} \cdot \mathbf{X} + \varphi)} + \sin{(-\mathbf{w}_{\ell} \cdot \mathbf{X} + \varphi)}~, \label{SUN1-mostrelevantdeformation}
\end{align}
where an arbitrary combination of the cosine and sine deformations \eqref{Vc}-\eqref{Vs} has been parameterized by an angle $\varphi$. The deformed action is then:
\begin{equation} \label{fullsuni1def}
    S[A] = \frac{1}{8 \pi}\int d^{2}x \, \gamma^{ab}  \partial_{a}\mathbf{X} \cdot \partial_{b}\mathbf{X} -i \frac{1}{2 \pi}\int dA^{i} \, \mathbf{w}_{i} \cdot \mathbf{X} + \lambda \int d^{2}x \,  V^{A_{\ell}}_{\mathbf{w}_{1}}(\varphi)~,
\end{equation}
where $\lambda$ is a dimensionful parameter that tends to $ \infty$ in the IR. As shown in detail in Appendix \ref{appendixmaximaandminima}, the global maxima and minima of this deformation at generic $\varphi$ are non-degenerate and belong to the set
\begin{equation} \label{critpoints}
    \mathbf{X}^{(j)} = 2 \pi \mathbf{w}_{j}~, \quad j=0, 1, \ldots, \ell~,
\end{equation}
where we define $\mathbf{w}_{0} \coloneqq 0$. Evaluated at these loci the potential takes the value
\begin{equation} \label{Aldeformationatextremapoint}
     V^{A_{\ell}}_{\mathbf{w}_{1}}(\varphi)\Big|_{\mathbf{X}^{(j)}} = (\ell + 1) \sin{\Big(\varphi- \frac{2 \pi j}{(\ell + 1)} \Big)}~.
\end{equation}
Thus for a typical $\varphi$ the potential is minimized by $ \mathbf{X}^{(j)} = 2 \pi \mathbf{w}_{j}$ for a particular $j$. As $\varphi$ is varied the minimizing value of $j$ jumps with isolated loci where there are two degenerate minima (See Figure \ref{fig:exampleplots}.)

Following the discussion around \eqref{bigreduced}, we can now use such relevant deformations to engineer any SPT transition for the symmetry group $PSU(M)$. Indeed, by appropriately choosing $\varphi$ we can arrange for the IR to be described by the trivially gapped phase around the unique vacuum $\mathbf{X}^{(j)} = 2 \pi \mathbf{w}_{j}$.  Thus, for $x<<0$ we deform the action by the potential $V^{A_{\ell}}_{\mathbf{w}_{1}}(\varphi_{-})$ with $\varphi_{-}$ chosen such that the vacuum is $\mathbf{X}\rightarrow 2\pi \mathbf{w}_{0}$. Similarly for $x>>0$ we deform the action by the potential $V^{A_{\ell}}_{\mathbf{w}_{1}}(\varphi_{+})$ with $\varphi_{+}$ chosen such that the vacuum is $\mathbf{X}\rightarrow 2\pi \mathbf{w}_{j}$.  Applying \eqref{bigreduced}, we see that this engineers a spatial transition of the type illustrated in Figure \ref{fig1}, where the SPT transitions corresponds to the $j$-th power of the generator in $H^{2}(PSU(M),U(1))\cong \mathbb{Z}_{M}$.  As this construction works for any $j$, any desired SPT transition can be achieved.

\subsubsection{Conformal Boundary Conditions from the Most-Relevant Perturbation} \label{subsubsection4.2.3}

In this subsection we apply the techniques of subsection \ref{subsection2.2} to the case of the $SU(M)_{1}$ WZW theory.  This allows us to develop a complete dictionary between the most relevant perturbations studied above and conformal boundary conditions.  Additionally, we reproduce the conclusions derived in the previous subsection.

Recall from equation \eqref{Cardy-energies} that in order to analyze the boundary conditions via Cardy's ansatz we need the modular data of the theory. The modular $S$ matrix in the case of the $SU(M)_{1}$ WZW theory is given by the $M$-th roots of unity:
\begin{equation}
    S_{ij} = \frac{\omega^{ij}}{\sqrt{M}}~, \hspace{1cm} \omega = e^{\frac{2 \pi i}{M}}~,
\end{equation}
where $i,j = 0, \ldots M-1$.  In terms of the non-abelian matrix valued field $g \in SU(M)$ (generalizing $g$ in \eqref{su2wzw}), $i=1$ and $i=M-1$ correspond to the highest weight representations with primaries $\mathrm{Tr}(g)$ and $\mathrm{Tr}(g^{\dagger})$ respectively.  In terms of the dual free boson variables $\mathbf{X}$ these correspond to the cosine and sine deformations defined in equation \eqref{SUN1-mostrelevantdeformation} as:
\begin{equation}
    C^{A_{\ell}}_{\mathbf{w}_{1}} \sim (\mathrm{Tr}(g^{\dagger}) + \mathrm{Tr}(g))/2~, \hspace{1cm} S^{A_{\ell}}_{\mathbf{w}_{1}} \sim (\mathrm{Tr}(g^{\dagger}) - \mathrm{Tr}(g))/2i~.
\end{equation}
This dictionary allows us to parameterize the $\varphi$ dependence of the potential as:
\begin{equation}
     V^{A_{\ell}}_{\mathbf{w}_{1}}(\varphi) = \cos{(\varphi)}S^{A_{\ell}}_{\mathbf{w}_{1}} + \sin{(\varphi)}C^{A_{\ell}}_{\mathbf{w}_{1}} =\frac{(\sin{(\varphi)} + i \cos{(\varphi)})}{2} \mathrm{Tr}(g) + \frac{(\sin{(\varphi)} - i \cos{(\varphi)})}{2} \mathrm{Tr}(g^{\dagger})~.
\end{equation}
We now apply the algorithm outlined in section \ref{subsection2.2} to determine the energies in  \eqref{Cardy-energies}:
\begin{equation} \label{minimizedsun1cardyenergies}
    E_{j} \propto - \Big[-\lambda \sin{\Big(\varphi-\frac{2 \pi j}{M} \Big)} \Big]^{\frac{2M}{M+1}}~,
\end{equation}
where $\lambda \rightarrow \infty$ in the IR, and $j$ on the left-hand side indicates the Cardy boundary state corresponding to the primary labelled by $\mathbf{w}_{j}$. 

According to the ansatz of \cite{Cardy:2017ufe}, the minimum among this set of energies (for finite but very large $\lambda$) determines the corresponding symmetry-preserving boundary condition. This happens when $\sin{\big( \varphi-\frac{2 \pi j}{M} \big)}$ attains its minimum value among all values of $j$. Note that this coincides exactly with the minimization problem for the potential 
$V^{A_{\ell}}_{\mathbf{w}_{1}}(\varphi)$ discussed around \eqref{Aldeformationatextremapoint}.  Therefore, we conclude that the relevant deformation $V^{A_{\ell}}_{\mathbf{w}_{1}}(\varphi)$ with $\varphi$ in the range such that the IR is $\mathbf{X}\rightarrow 2\pi \mathbf{w}_{j}$, engineers the conformal boundary conditions associated to the primary labelled by $\mathbf{w}_{j}$ when activated in half of spacetime.

\section{$\mathbb{Z}_{2}\times \mathbb{Z}_{2}$ SPT Transition via Gauged Ising Models} \label{section5}

In this section we consider an example of an SPT transition associated to the discrete global symmetry group $G\cong\mathbb{Z}_{2}\times \mathbb{Z}_{2}$.  The relevant group cohomology characterizing SPTs is:
\begin{equation*}
   H^{2}( \mathbb{Z}_{2}\times \mathbb{Z}_{2},U(1))\cong \mathbb{Z}_{2}~.
\end{equation*}
Thus, we aim to describe a transition between a trivial SPT and the unique non-trivial SPT.  The latter can be characterized in terms of two $\mathbb{Z}_{2}$ background gauge fields $A_{1}, A_{2}$ as 
\begin{equation}\label{z2z2spt}
   Z_{\text{SPT}}= \exp\left(i\pi \int_{X} A_{1}\cup A_{2}\right)~.
\end{equation}

\subsection{Bulk Analysis of Relevant Deformations} \label{Z2Z2-subsection1}

The critical point mediating the transition which generates \eqref{z2z2spt} will be two copies of the Ising model appropriately gauged.  Specifically, each Ising model has a $\mathbb{Z}_{2}$ global symmetry and we gauge these with dynamical $\mathbb{Z}_{2}$ gauge fields $a_{i}$ together with a non-trivial Dijkgraaf-Witten term \cite{Dijkgraaf:1989pz} coupling $a_{1}$ and $a_{2}$.  The resulting model has a dual (or orbifold) symmetry $G \cong \mathbb{Z}_{2}\times \mathbb{Z}_{2}$ with background fields $A_{1}, A_{2}$.  The action is thus:  
\begin{equation} \label{XY}
    S= \sum_{i=1}^{2} \bigg[ \int (\mathcal{D}_{a_{i}}\sigma_{i})^{2} + \sigma_{i}^{4} + i \pi \int A_{1} \cup a_{i} \bigg] + i \pi \bigg[ \int a_{1} \cup a_{2} + \int a_{2} \cup A_{2} \bigg]~.
\end{equation}
The model \eqref{XY} enjoys several dualities which are useful below.  First, using the relationship between gauged Ising and a Majorana fermion, one can recast the above as a theory of two Majorana fermions $\chi_{i}$ (one Dirac) coupled to a $\mathbb{Z}_{2}$ gauge field $c$ with appropriate action (see e.g. \cite{Karch:2019lnn}):
\begin{equation}\label{fermframe}
    S=\int i \overline{\chi}_{1} \slashed{D}_{(c+A_{1}) \cdot \rho} \chi_{1} + i \overline{\chi}_{2} \slashed{D}_{(c+A_{1}+A_{2}) \cdot \rho} \chi_{2} + i \pi \mathrm{Arf}[s \cdot \rho]~,
\end{equation}
where $\rho$ stands for a spin structure and Arf is the quadratic action defined by the Arf invariant. (Note that since $\rho$ is effectively summed over, the resulting model is bosonic as is manifest in \eqref{XY}).

The description in terms of a Dirac fermion $\chi_{1}+i\chi_{2}$ with its continuous $U(1)$ global rotation symmetry also makes manifest that the model \eqref{XY}-\eqref{fermframe} enjoys yet another dual description via bosonization as a compact scalar at radius $R=2$. We will make use of this description below.

In order to see the SPT transition we use the Ising presentation \eqref{XY}.  We deform the model by the relevant quadratic scalar operators:
\begin{equation}\label{isingpot}
    \delta S= \int d^{2}x~\alpha \sigma_{1}^{2} + \beta \sigma_{2}^{2}~,
\end{equation}
where $\alpha$ and $\beta$ are coefficients whose magnitudes grow in the IR.  The vacuum structure now depends on the sign of each of these coefficients.

\begin{itemize}
    \item $\alpha>0$ and $\beta>0$.  At low energies the Ising fields $\sigma_{i}$ are frozen to zero.  The resulting action for the discrete gauge fields $a_{i}$ is:
    \begin{equation}
        S\rightarrow i\pi \int a_{1}\cup (A_{1}+a_{2})+i\pi \int a_{2}\cup (A_{1}+A_{2})~.
    \end{equation}
    The path integral over $a_{1}$ now imposes $a_{2}=A_{1}$ so that all dynamical variables are fixed and we are left with a non-trivial SPT phase of the form \eqref{z2z2spt}:
\begin{equation}
    S\rightarrow i\pi \int A_{1}\cup A_{2}~,
\end{equation}
where above, we have assumed that our spacetime manifold is orientable (equivalently we do not probe time-reversal symmetry) to simplify $A_{1}\cup A_{1}=A_{1}\cup w_{1}=0$, with $w_{1}$ the first Stiefel-Whitney class of spacetime.  

\item $\alpha<0$ and $\beta<0$.  Now the potential \eqref{isingpot} condenses the Ising fields at a non-zero value.  The dynamical $\mathbb{Z}_{2}\times \mathbb{Z}_{2}$ gauge symmetry permutes these values leading to a unique ground state.  Moreover, the gauging also ensures that in the IR we have effectively $a_{i}=0$.  Thus in this regime we end up with a trivially gapped theory with trivial SPT action.

\item $\alpha>0$ and $\beta<0$ or $\alpha<0$ and $\beta>0$.  In this regime one of the Ising scalars is frozen to the origin and one condenses.  The latter trivializes the associated gauge field but the former does not, leading to an IR that has two ground states (labelled by the local operators dual to the non-trivial $a_{i}$) and thus spontaneously breaks $\mathbb{Z}_{2}\times \mathbb{Z}_{2}\rightarrow \mathbb{Z}_{2}.$
\end{itemize}

\begin{figure}[t]
        \centering
        \includegraphics[width=0.5 \textwidth]{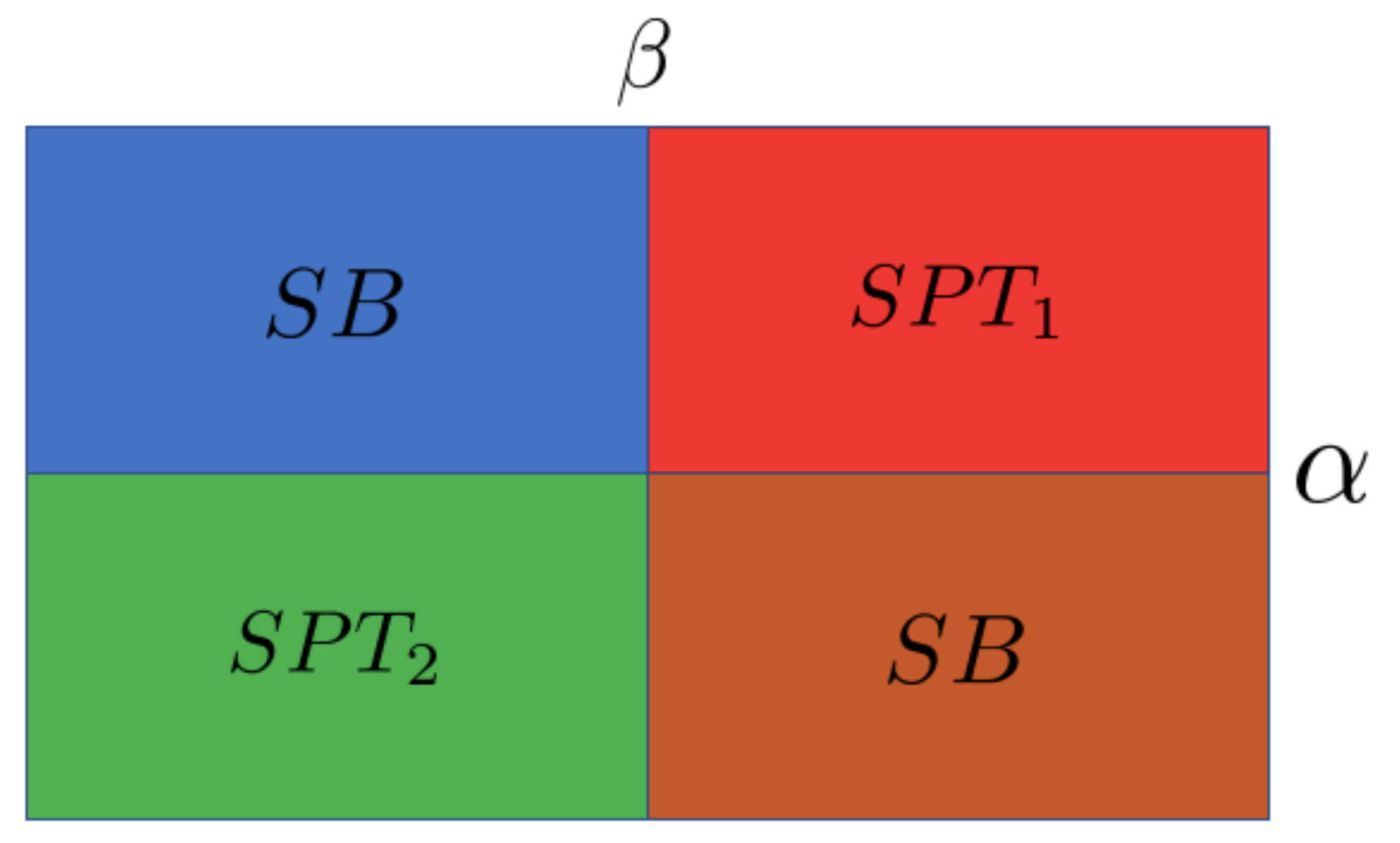} % second figure itself
        \caption{Phase diagram for the $\mathbb{Z}_{2} \times \mathbb{Z}_{2}$ transition studied in section \ref{section5}.  In the first and third quadrant, where $\alpha$ and $\beta $ in the potential \eqref{isingpot} have the same sign, we obtain a trivially gapped IR with distinct $\mathbb{Z}_{2}\times Z_{2}$ SPTs.  In the second an fourth quadrant, the symmetry is spontaneously broken $\mathbb{Z}_{2}\times \mathbb{Z}_{2}\rightarrow \mathbb{Z}_{2}.$}\label{fig6}
\end{figure}

The phase diagram resulting from our analysis agrees with that of \cite{Tsui:2017ryj} and is summarized in Figure \ref{fig6}.  Focusing on the regimes where the symmetry is not spontaneously broken, we see that we can generate a $\mathbb{Z}_{2}\times \mathbb{Z}_{2}$ SPT transition by, for instance, taking $\alpha=\beta$  from $-\infty$ to $+\infty.$

\subsection{$\mathbb{Z}_{2} \times \mathbb{Z}_{2}$ Transition in the Boundary CFT} \label{Z2Z2-subsection2}

We now study the transition described in the previous subsection from the perspective of boundary conformal field theory. For this analysis, it is helpful to use the presentation of \eqref{XY}-\eqref{fermframe} in terms of a compact scalar $X$ at radius $R=2$. In this duality frame the two $\mathbb{Z}_{2}$ symmetries act as:
\begin{equation}
    A_{1}: X \longrightarrow X + \pi~, \hspace{.2in} A_{2}: X \longrightarrow -X~. 
\end{equation}
All conformal boundary conditions at $c=1$ for the circle branch are known \cite{Gaberdiel:2001zq,Janik:2001hb}, and a summary can be found in appendix \ref{appendixirrationaltheories}. It is straightforward to check that the boundary conditions that preserve the previous symmetries are given by the Neumann conditions (see relatedly \cite{Cho:2016xjw}):
\begin{align}
    | N(0) \} &=   \sum_{J=0}^{\infty} (-1)^{J} | J \rrangle + \sum_{w \in \mathbb{Z}/\{0\}} | (0,w)\rrangle~, \label{Neumann0} \\ 
    | N(\pi) \} &=   \sum_{J=0}^{\infty} (-1)^{J} | J \rrangle + \sum_{w \in \mathbb{Z}/\{0\}} (-1)^{w} | (0,w)\rrangle~,  \label{Neumannpi}
\end{align}
and we can read off explicitly the degeneracies and energies of the cylinder theory by computing the cylinder partition function:
\begin{equation}
    Z_{N(0)N(\pi)} = \frac{1}{\eta(q)} \sum_{k \in \mathbb{Z}} q^{\frac{1}{8}(2k+1)^{2}} = \frac{1}{\eta(q)} \big( 2q^{1/8} + 2q^{9/8} + \cdots \big)~.
\end{equation}
In particular, notice that all states are indeed two-fold degenerate which is consistent with the fact that each energy level must furnish a nontrivial projective representation of the $\mathbb{Z}_{2}\times \mathbb{Z}_{2}$ global symmetry. 

In order to evaluate a bound on the central charge along the lines of section \ref{section3}, let us note that the cylinder partition function is not written in terms of a finite sum of characters so we cannot apply \eqref{inequalityfromCardyCond} directly. Instead, as pointed out in Appendix \ref{appendixirrationaltheories}, since the coefficients of the Ishibashi states in \eqref{Neumann0}-\eqref{Neumannpi} are phases (or more generally for the boundary states \eqref{Dirichlet}-\eqref{Neumann}) their magnitude is at most one. Thus the implied bound on the cylinder partition function in terms of the torus partition function is simply $Z_{AB}(\beta) \leq Z_{T^{2}}(2\beta)$.

It is now straightforward to proceed as in section \ref{section3}. Specifically, one can use \eqref{bndiag} with the replacement: $ \Big( \sum_{j} n^{j}_{AB} \Big) \rightarrow 1$. For the choice of threshold $\Delta_{BH}$ in \eqref{simpthresh} the only light primary corresponds to the identity operator so $N_{\Delta_{BH}} = 1$. The central charge and the conformal weight associated to the ground state are:
\begin{equation}
    c= 1~, \hspace{.2in} h_{N(0)N(\pi),\text{min}} = 1/8~.
\end{equation}
The result of the minimization then gives:
\begin{equation}
    d_{N(0)N(\pi)} = 2 < 8.53775 \ldots ,
\end{equation}
where the right hand side is the result of the minimization.

\section*{Acknowledgements}

We thank N. Afkhami-Jeddi, and Y. Wang for discussion and collaboration at an early stage of this project. CC and DGS acknowledge support from the Simons Collaboration on Global Categorical Symmetries, and the US Department of Energy DE-SC0009924. 

\appendix

\section{First Order Transitions} \label{appendixfirstordertransitions}

In this appendix we discuss first order transitions between SPTs mediated by a simple topological quantum field theory.  We will outline the main points by considering a $\mathbb{Z}_{N}$ gauge theory in two dimensions. 

A continuum action for such a theory may be written as \cite{Banks:2010zn,Maldacena:2001ss,Kapustin:2014gua}:
\begin{equation} \label{BF-theory}
    S_{BF} = \frac{i N}{2\pi}\int \varphi \wedge d a~,
\end{equation}
where $\varphi$ is a periodic scalar $\varphi \sim \varphi + 2 \pi$ and $a$ is a $\mathbb{Z}_{N}$ gauge field.  The local operators in the theory are given by:
\begin{equation}\label{topops}
    \mathcal{O}_{k} = \exp{( i \, k \, \varphi )}~,
\end{equation}
with $k$ an integer mod $N$: $k \sim k + N$.  This theory has $N$ ground states which we denote as $| i \rangle$ with $i=0,\ldots,N-1$.  In these states the topological local operators in \eqref{topops} have expectation values: 
\begin{equation} \label{expectation-value}
    \langle r | \mathcal{O}_{k} | s \rangle = \delta_{r,s} \exp{ \Big( 2 \pi i \, \frac{ s k }{N} \Big)}~ \longleftrightarrow  \langle s | \varphi | s \rangle =\frac{2\pi s}{N}~.
\end{equation}

Let us deform the action by the local operators.  Since these operators are topological they have dimension zero and are hence relevant.  Thus they modify the relative energies of the ground states.  (See \cite{Hellerman:2006zs, Cherman:2021nox} and references therein for a more detailed analysis.)  For instance, consider a deformation of the form
\begin{equation}
    \Delta S \sim \lambda  \int d^{2}x \big( \exp(i\chi)\mathcal{O}_{k} +\exp(-i\chi) \mathcal{O}^{\dagger}_{k} \big)\sim \lambda \int d^{2}x \left(\cos(\chi+k\varphi)\right)~, 
\end{equation}
where above $\lambda>0$ and $\chi$ is an arbitrary angle.  Clearly by appropriately choosing the angle $\chi$ and $k$ we can arrange for any state $|s\rangle$ to be the non-degenerate ground state.  Below we will assume that we have deformed the theory such that for $x\ll0$ there is a non-degenerate ground state $|0\rangle$ and for $x\gg0$ there is a non-degenerate ground state $|1\rangle.$

 We can use the deformation constructed above to generate SPT transitions for any SPT phase characterized by an action of order $N$.  In the notation introduced around \eqref{SPTacttrans} we are thus assuming that the SPT action $\omega_{+-}(A)$ obeys that $N\omega_{+-}(A)\sim 0$.  To do this, we first couple the $\mathbb{Z}_{N}$ gauge theory to its $\mathbb{Z}_{N}^{(1)}$ one-form symmetry \cite{Gaiotto:2014kfa} with two-form background field $B\in H^{2}(X,\mathbb{Z}_{N})$:
\begin{equation}\label{bact}
    S[B] = \frac{i N}{2\pi}\int \varphi \wedge d a + i \int \varphi \wedge B ~.
\end{equation}
Activating the deformation described above implies that this TQFT can mediate a transition between the general one-form SPT of order $N$:
\begin{equation}\label{1formspttrans}
    \frac{Z_{IR+}[B]}{Z_{IR-}[B]}=\exp\left(\frac{2\pi i}{N}\int B \right)~.
\end{equation}
We now utilize the one-form symmetry to couple the ordinary background gauge fields by substituting $B=\omega_{+-}(A)$ into the action \eqref{bact}.  (This is often called ``symmetry fractionalization" see e.g.\ \cite{Benini:2018reh, Brennan:2022tyl, Delmastro:2022pfo} for more examples and recent analysis).  Then the construction above gives us a transition between any SPTs of order $N$ where the transition is first order:
\begin{equation}\label{0formspttransgen}
    \frac{Z_{IR+}[A]}{Z_{IR-}[A]}=\exp\left(\frac{2\pi i}{N}\int \omega_{+-}(A) \right)~.
\end{equation}

\section{Sine and Cosine Deformations}
\label{appendixoncocycles}

In this appendix we provide the details of how the sine and cosine deformations \eqref{Vc} and \eqref{Vs} are preserved under an $A_{\ell}$ algebra. Let us recall that in order to generate the appropriate affine symmetry algebra, the naive holomorphic vertex operator $\tilde{V}_{\mathbf{r}_{i}}(z) = e^{i \mathbf{r}_{i} \cdot \mathbf{X}^{L}(z)}$ must be dressed with a correction factor $c_{\mathbf{r}_{i}}$ that only depends on the momentum part of the free boson zero modes: $V_{\mathbf{r}_{i}} = c_{\mathbf{r}_{i}}(\hat{p}_{L}) \tilde{V}_{\mathbf{r}_{i}}$. Below we will find the combination of zero modes generating the $A_{\ell}$ algebra which annihilates $C^{A_{\ell}}_{\mathbf{w}_{1}}$ and $S^{A_{\ell}}_{\mathbf{w}_{1}}$, and we will further see how the extremization problem considered in the main text can be reduced to the case where the left/right momentum modes are taken to zero. The summary below is based on \cite{DiFrancesco:639405} (see also \cite{green_schwarz_witten_2012}).

Let us first list a few properties of $c_{\mathbf{r}_{i}}(\hat{p}_{L})$. When $c_{\mathbf{r}}(\hat{p}_{L})$ passes over a vertex, its argument is shifted:
\begin{equation}
    e^{i \mathbf{r}\cdot \mathbf{X}^{L}(z)} c_{\mathbf{s}}(\hat{p}_{L}) = c_{\mathbf{s}}(\hat{p}_{L} - \mathbf{r}) e^{i \mathbf{r}\cdot \mathbf{X}^{L}(z)}~.
\end{equation}
To recover the appropriate algebra from the OPE we need
\begin{equation} \label{condition1}
    c_{\mathbf{r}}(\hat{p}_{L}) c_{\mathbf{s}}(\hat{p}_{L}-\mathbf{r}) = (-1)^{(\mathbf{r},\mathbf{s})} c_{\mathbf{s}}(\hat{p}_{L}) c_{\mathbf{r}}(\hat{p}_{L}-\mathbf{s})~,
\end{equation}
and to further obtain a closed algebra we require
\begin{equation} \label{condition2}
    c_{\mathbf{r}}(\hat{p}_{L}) c_{\mathbf{s}}(\hat{p}_{L}-\mathbf{r}) = \epsilon(\mathbf{r},\mathbf{s}) c_{\mathbf{r} + \mathbf{s}}(\hat{p}_{L})~,
\end{equation}
where $\epsilon(\mathbf{r},\mathbf{s}) = \pm 1$. An explicit construction that fulfills these conditions can be obtained in the following way. Let $\mathbf{r}_{i}$ be the simple roots of the algebra, and expand
\begin{equation}
    \mathbf{r} = \sum n_{i} \mathbf{r}_{i}~, \quad \mathbf{s} = \sum m_{i} \mathbf{r}_{i}~, \quad n_{i},m_{i} \in \mathbb{Z}~.
\end{equation}
Now introduce the product
\begin{equation} \label{starproduct}
    \mathbf{r} \ast \mathbf{s} = \sum_{i>j} n_{i} m_{j} (\mathbf{r}_{i} ~, \mathbf{r}_{j})~.
\end{equation}
Then, it is straightforward to check that
\begin{equation} \label{correctionfactordef}
    c_{\mathbf{r}}(\hat{p}_{L}) = (-1)^{\hat{p}_{L} \ast \mathbf{r}}
\end{equation}
fulfills the equations \eqref{condition1} and \eqref{condition2}, with $\epsilon(\mathbf{r}, \mathbf{s}) = (-1)^{\mathbf{r} \ast \mathbf{s}}$. Once this choice has been made at the level of the algebra, we have that for other representations of the current algebra, the correction factor is given by a shift on the zero-mode momentum operator: $c_{\mathbf{w}} = (-1)^{(\hat{p}_{L}-\tilde{\mathbf{w}})\ast \mathbf{w}}$, where $\tilde{\mathbf{w}}$ is the highest-weight of the representation to which $\mathbf{w}$ belongs.

An important point is that the previous construction is not unique: any solution of the fundamental equations is equivalent. In particular, we can make different choices in the holomorphic and antiholomorphic sectors of our CFT since they generate different, commuting algebras. Let us use \eqref{correctionfactordef} on the holomorphic side, and $\overline{c}_{\mathbf{r}}(\hat{p}_{R}) = (-1)^{-\hat{p}_{R} \ast \mathbf{r}}$ on the antiholomorphic side. This will make several cancellations below manifest, and it is straightforward to see that it fulfills \eqref{condition1} and \eqref{condition2}. The corresponding correction factors on the antiholomorphic side will then appear with a respective minus sign.

Let us now exhibit the  annihilation of the cosine deformation in $SU(M)_{1}$ using the above construction. If we take the $(i+1)$-th summand of the deformation in \eqref{Vc} with appropriate correction factors attached, and apply the holomorphic raising operator associated to the simple root $\mathbf{r}_{i}$ to it:
\begin{align}
    (-1)^{\hat{p}_{L} \ast \mathbf{r}_{i}} e^{i \mathbf{r}_{i} \cdot \mathbf{X}^{L}(z)}\Bigg( & (-1)^{(\hat{p}_{L} - \mathbf{w}_{1})\ast (\mathbf{w}_{i+1}- \mathbf{w}_{i})} (-1)^{-(\hat{p}_{R} - \mathbf{w}_{1})\ast (\mathbf{w}_{i+1}- \mathbf{w}_{i})} e^{i(\mathbf{w_{i+1}}-\mathbf{w}_{i})\cdot (\mathbf{X}^{L}(\omega) + \mathbf{X}^{R}(\overline{\omega}))} \nonumber \\ &+ (-1)^{(\hat{p}_{L} - \mathbf{w}_{\ell})\ast (\mathbf{w}_{i}- \mathbf{w}_{i+1})} (-1)^{-(\hat{p}_{R} - \mathbf{w}_{\ell})\ast (\mathbf{w}_{i}- \mathbf{w}_{i+1})} e^{i(\mathbf{w_{i}}-\mathbf{w}_{i+1})\cdot (\mathbf{X}^{L}(\omega) + \mathbf{X}^{R}(\overline{\omega}))} \Bigg)/2 \nonumber
\end{align}
\begin{align}
    & \sim \frac{(-1)^{\hat{p}_{L} \ast \mathbf{r}_{i}} (-1)^{(\hat{p}_{L} - \mathbf{r}_{i})\ast (\mathbf{w}_{i+1}- \mathbf{w}_{i})} (-1)^{-\hat{p}_{R} \ast (\mathbf{w}_{i+1}- \mathbf{w}_{i})} e^{i(\mathbf{r}_{i} + \mathbf{w}_{i+1} - \mathbf{w}_{i})\cdot \mathbf{X}^L(\omega) + (\mathbf{w}_{i+1} - \mathbf{w}_{i})\cdot \mathbf{X}^{R}(\overline{\omega})}}{2(z-\omega)} \nonumber \\
    & = \frac{ (-1)^{\hat{p}_{L}\ast (\mathbf{w}_{i}- \mathbf{w}_{i-1})} (-1)^{-\hat{p}_{R} \ast (\mathbf{w}_{i+1}- \mathbf{w}_{i})} (-1)^{-\mathbf{r}_{i}\ast(\mathbf{w}_{i+1}-\mathbf{w}_{i})} e^{i(\mathbf{w}_{i} - \mathbf{w}_{i-1})\cdot \mathbf{X}^L(\omega) + (\mathbf{w}_{i+1} - \mathbf{w}_{i})\cdot \mathbf{X}^{R}(\overline{\omega})}}{2(z-\omega)}~. \label{OPE1}
\end{align}
Notice how only the first summand of the cosine deformation above contributes to the OPE. Consider now the OPE involving the $i$-th summand of the deformation with the antiholomorphic lowering operator:
\begin{align}
    (-1)^{\hat{p}_{R} \ast \mathbf{r}_{i}} e^{- i \mathbf{r}_{i} \cdot \mathbf{X}^{R}(\overline{z})}\Bigg( & (-1)^{(\hat{p}_{L} - \mathbf{w}_{1})\ast (\mathbf{w}_{i}- \mathbf{w}_{i-1})} (-1)^{-(\hat{p}_{R} - \mathbf{w}_{1})\ast (\mathbf{w}_{i}- \mathbf{w}_{i-1})} e^{i(\mathbf{w_{i}}-\mathbf{w}_{i-1})\cdot (\mathbf{X}^{L}(\omega) + \mathbf{X}^{R}(\overline{\omega}))} \nonumber \\ &+ (-1)^{(\hat{p}_{L} - \mathbf{w}_{\ell})\ast (\mathbf{w}_{i-1}- \mathbf{w}_{i})} (-1)^{-(\hat{p}_{R} - \mathbf{w}_{\ell})\ast (\mathbf{w}_{i-1}- \mathbf{w}_{i})} e^{i(\mathbf{w_{i-1}}-\mathbf{w}_{i})\cdot (\mathbf{X}^{L}(\omega) + \mathbf{X}^{R}(\overline{\omega}))} \Bigg)/2 \nonumber
\end{align}
\begin{align}
    & \sim \frac{(-1)^{\hat{p}_{R} \ast \mathbf{r}_{i}} (-1)^{\hat{p}_{L}\ast (\mathbf{w}_{i}- \mathbf{w}_{i-1})} (-1)^{ -(\hat{p}_{R} + \mathbf{r}_{i}) \ast (\mathbf{w}_{i} - \mathbf{w}_{i-1})} e^{i(\mathbf{w}_{i} - \mathbf{w}_{i-1})\cdot \mathbf{X}^L(\omega) + (-\mathbf{r}_{i}+\mathbf{w}_{i} - \mathbf{w}_{i-1})\cdot \mathbf{X}^{R}(\overline{\omega})}}{2(\overline{z}-\overline{\omega})} \nonumber \\
    & = \frac{  (-1)^{\hat{p}_{L}\ast (\mathbf{w}_{i}- \mathbf{w}_{i-1})} (-1)^{ \hat{p}_{R} \ast (\mathbf{w}_{i}-\mathbf{w}_{i+1})} (-1)^{-\mathbf{r}_{i} \ast (\mathbf{w}_{i} - \mathbf{w}_{i-1})} e^{i(\mathbf{w}_{i} - \mathbf{w}_{i-1})\cdot \mathbf{X}^L(\omega) + (\mathbf{w}_{i+1} - \mathbf{w}_{i})\cdot \mathbf{X}^{R}(\overline{\omega})}}{2(\overline{z}-\overline{\omega})}~. \label{OPE2}
\end{align}
The numerators of \eqref{OPE1} and \eqref{OPE2} coincide then up to phases $(-1)^{-\mathbf{r}_{i}\ast(\mathbf{w}_{i+1}-\mathbf{w}_{i})}$ and $(-1)^{-\mathbf{r}_{i} \ast (\mathbf{w}_{i} - \mathbf{w}_{i-1})}$. It is straightforward to see from the definition \eqref{starproduct}, and from the fact that $\mathbf{r}_{i} = - \mathbf{w}_{i-1} + 2 \mathbf{w}_{i} - \mathbf{w}_{i+1}$ for $su(M)$ that:
\begin{equation}
    (-1)^{\mathbf{r}_{i} \ast \mathbf{r}_{i}} = 1 = (-1)^{\mathbf{r}_{i} \ast (- \mathbf{w}_{i-1} + 2 \mathbf{w}_{i} - \mathbf{w}_{i+1})} \Longrightarrow (-1)^{-\mathbf{r}_{i}\ast(\mathbf{w}_{i+1}-\mathbf{w}_{i})} = (-1)^{-\mathbf{r}_{i} \ast (\mathbf{w}_{i} - \mathbf{w}_{i-1})}~.
\end{equation}
We see that the zero-modes of the holomorphic and antiholomorphic currents in \eqref{OPE1} and \eqref{OPE2} indeed annihilate the cosine deformation. Notice further that since in both cases only the first summands contribute in the OPE it follows that the same conclusion holds for the sine deformation. In general, if we call the modes of the $c_{i}(\hat{p}_{L}) e^{i \mathbf{r}_{i} \cdot \mathbf{X}^{L}(z)}$ current $e^{i}_{n}$, and the modes of the $\overline{c}_{-i}(\hat{p}_{R}) e^{- i \mathbf{r}_{i} \cdot \mathbf{X}^{R}(\overline{z})}$ current $\overline{f}^{i}_{n}$, we find that the combination $e^{i}_{0} - \overline{f}^{i}_{0}$ annihilates the cosine and sine deformations. With a similar computation we can check that, if we call $f^{i}_{n}$ the modes of the current $c_{-i}(\hat{p}_{L}) e^{-i \mathbf{r}_{i} \cdot \mathbf{X}^{L}(z)}$, and $\overline{e}^{i}_{n}$ the modes of the current $\overline{c}_{i}(\hat{p}_{R}) e^{ i \mathbf{r}_{i} \cdot \mathbf{X}^{R}(\overline{z})}$, that the combination $f^{i}_{0} - \overline{e}^{i}_{0}$ also annihilates the cosine and sine deformations. Meanwhile, the fact that the combination $h^{i}_{0}-\overline{h}^{i}_{0}$ of holomorphic and antiholomorphic Cartan subalgebra zero-modes also annihilates the sine and cosine deformation is clear. It is also straightforward to check that the previous combinations of zero modes fulfill the zero-mode algebra of the $SU(2)$ subalgebras of $SU(M)$, and since the action of non-simple roots can be obtained from commutators of simple root generators, we find that the cosine and sine deformation are annihilated by the full algebra as required.

Finally, let us comment on why the ground state analysis performed in the main text is insensitive to the presence of these cocycle dressings.  In principle, the relevant sine and cosine deformations of the action should appear with insertions of $p_{L}$ and $p_{R}$. Moreover, in the path integral we are instructed to sum over these discrete momentum labels, organized as vectors with quantized entries living in the same conjugacy class of the weight lattice (i.e.\ $p_{L}-p_{R}$ is in the root lattice). However, notice that only the case where $p_{L}$ and $p_{R}$ are both zero leads to a vanishing kinetic term.  Thus, in analyzing the vacua and SPT phases as done in subsections \ref{subsection4.2.2}, and appendices \ref{appendixmaximaandminima} and \ref{Other-Simply-Laced-Algebras} we are justified in truncated to the sector of zero momentum.

\section{$SU(M)_{1}$ Deformation: Maxima and Minima of $V_{\mathbf{w}_{1}}(\varphi)$} \label{appendixmaximaandminima}

In this appendix we show analytically that the sine-plus-cosine deformation $V_{\mathbf{w}_{1}}(\varphi)$ for $SU(M)_{1}$ has minima and maxima at $\mathbf{X} = 2 \pi \mathbf{w}_{i}$. We proceed directly, by checking the conditions required for critical points, and analyzing the positive-definiteness of the Hessian to establish the character of local maxima or minima for the $\mathbf{X} = 2 \pi \mathbf{w}_{i}$ critical points.

Up to an overall constant, a general linear combination of sines and cosines can be written as a sine with an additional phase $\varphi.$  Thus we write the deformation as:
\begin{equation} \label{deformation-appendix}
    V_{\mathbf{w}_{1}}(\varphi) = \sin{(\mathbf{w}_{1} \cdot \mathbf{X} + \varphi)} + \sum_{i=1}^{\ell-1} \sin{ (\mathbf{w}_{i+1} \cdot \mathbf{X} - \mathbf{w}_{i} \cdot \mathbf{X} + \varphi)} + \sin{(-\mathbf{w}_{\ell} \cdot \mathbf{X} + \varphi)}~,
\end{equation}
where $\ell = M-1$ is the rank of the group $SU(M)$.

In order to proceed let us work with the variables $Y_{i} = \mathbf{w}_{i} \cdot \mathbf{X}.$ The conditions obeyed by a critical point are
\begin{eqnarray}
    \frac{\partial V_{\mathbf{w}_{1}}(\varphi)}{\partial Y_{1}}& =& -2\sin{\Big(\frac{Y_{2} + 2\varphi}{2}\Big)} \sin{\Big(\frac{2Y_{1}-Y_{2}}{2}\Big)} = 0~, \label{crit1} \\ \frac{\partial V_{\mathbf{w}_{1}}(\varphi)}{\partial Y_{i}}&=& -2\sin{\Big(\frac{Y_{i+1} - Y_{i-1} + 2\varphi}{2}\Big)} \sin{\Big(\frac{2Y_{i}-Y_{i-1}-Y_{i+1}}{2}\Big)} = 0~, \label{crit2}\\ \frac{\partial V_{\mathbf{w}_{1}}(\varphi)}{\partial Y_{\ell}} &=& -2\sin{\Big(\frac{-Y_{\ell - 1} + 2 \varphi}{2}\Big)} \sin{\Big(\frac{2Y_{\ell}-Y_{\ell-1}}{2}\Big)} = 0~.  \label{crit3}
\end{eqnarray}
Let us concentrate on the set of critical points associated to the vanishing of the right factors. As we will see the global maximum (minimum) is actually contained in this set. Then, the critical points in this set must satisfy
\begin{equation} \label{criticalitycondition}
\begin{pmatrix}
2 & -1 & 0 & 0 & \ldots & 0 & 0 \\
-1 & 2 & -1 & 0 & \ldots & 0 & 0 \\
0 & -1 & 2 & -1 & \ldots & 0 & 0 \\
. & . & . & . & \ldots & . & . \\
0 & 0 & 0 & 0 & \ldots & 2 & -1 \\
0 & 0 & 0 & 0 & \ldots & -1 & 2 \\
\end{pmatrix}
\begin{pmatrix}
Y_{1} \\ Y_{2} \\ Y_{3} \\ \vdots \\ Y_{\ell-1} \\ Y_{\ell}
\end{pmatrix} = 2 \pi\begin{pmatrix}
n_{1} \\ n_{2} \\ n_{3} \\ \vdots \\ n_{\ell-1} \\ n_{\ell}
\end{pmatrix}
\end{equation}
for some set of integers $n_{i}$. Noticing that the matrix is the Cartan matrix of $A_{\ell}$ relating roots and weights, we can recast this condition as $\mathbf{X} \cdot \mathbf{r}_{i} = 2 \pi n_{i}$. Expanding $\mathbf{X}$ in terms of the fundamental weights as $\mathbf{X} = 2 \pi \sum_{j} a_{j} \mathbf{w}_{j}$ we see that the condition for criticality is solved if and only if $a_{i} = n_{i}$. Clearly, not all these critical points are independent as we have to take the periodicity of the space in the root lattice into account. From the Cartan matrix we can see that $\mathbf{w}_{1} + \mathbf{w}_{i} \sim \mathbf{w}_{i+1}$ for $1 \leq i < \ell$ and $\mathbf{w}_{1} + \mathbf{w}_{\ell} \sim 0$, where $\sim$ stands for equality up to roots. Using this is easy to prove that $n \mathbf{w}_{1} \sim \mathbf{w}_{n}$ for $1 \leq n \leq \ell$, and $(\ell + 1) \mathbf{w}_{1} \sim 0$. Thus, any critical point $\mathbf{X} = 2 \pi \sum_{j} n_{j} \mathbf{w}_{j}$ can indeed be identified to either 0 or one of the independent critical points $\mathbf{X}^{(j)} = 2 \pi \mathbf{w}_{j}$. 

We now show that these critical points indeed correspond to strict maxima and minima. The Hessian is computed from 
\begin{eqnarray}
       \frac{\partial^{2}V_{\mathbf{w}_{1}}(\varphi)}{\partial Y_{1}^{2}} &=& -\sin{(Y_{1} + \varphi)} - \sin{(Y_{2}-Y_{1}+ \varphi)}~, \label{hessian1} \\ \frac{\partial^{2}V_{\mathbf{w}_{1}}(\varphi)}{\partial Y_{i}^{2}} &=& -\sin{(Y_{i} - Y_{i-1} + \varphi)} - \sin{(Y_{i+1} - Y_{i} + \varphi)}~, \label{hessian2}\\ \frac{\partial^{2}V_{\mathbf{w}_{1}}(\varphi)}{\partial Y_{\ell}^{2}} &=& - \sin{(Y_{\ell} - Y_{\ell - 1} + \varphi)} -\sin{(-Y_{\ell} + \varphi)}~,  \label{hessian3} \\ \frac{\partial^{2}V_{\mathbf{w}_{1}}(\varphi)}{\partial Y_{i+1} \partial Y_{i}} &=& \sin{(Y_{i+1} - Y_{i} + \varphi)}~,\label{hessian4} 
\end{eqnarray}
with other second derivatives vanishing. Now, when $\mathbf{X}^{(j)} = 2 \pi \mathbf{w}_{j}$, $Y^{(j)}_{i} = 2 \pi (A^{-1})_{ij}$, where $(A^{-1})_{ij}$ is the corresponding inverse Cartan matrix
\begin{equation}
    (A^{-1})_{ij} = \mathrm{min}(i,j) - \frac{ij}{(\ell + 1)}~. \label{suninversecartan}
\end{equation}
So, evaluating \eqref{hessian1}-\eqref{hessian4} in these points, we can see that
\begin{align}
    \frac{\partial^{2} V_{\mathbf{w}_{1}}(\varphi)}{\partial Y_{i}^{2}}\bigg|_{Y^{(j)}} &=  -2\sin{\Big(-\frac{2 \pi j}{(\ell + 1)} + \varphi \Big)}~, \\ \frac{\partial^{2} V_{\mathbf{w}_{1}}(\varphi)}{\partial Y_{i+1} \partial Y_{i}}\bigg|_{Y^{(j)}} &=  \sin{\Big(-\frac{2 \pi j}{(\ell + 1)} + \varphi \Big)}~.
\end{align}
We observe the Hessian corresponds to a tridiagonal, Toeplitz-type matrix. This is, a $n \times n$ matrix of the form
\begin{equation} \label{toeplitz}
\begin{pmatrix}
a & c & 0 & 0 & \ldots & 0 & 0 \\
b & a & c & 0 & \ldots & 0 & 0 \\
0 & b & a & c & \ldots & 0 & 0 \\
. & . & . & . & \ldots & . & . \\
0 & 0 & 0 & 0 & \ldots & a & c \\
0 & 0 & 0 & 0 & \ldots & b & a \\
\end{pmatrix}~,
\end{equation}
%\begin{equation} \label{toeplitz}
%\begin{pmatrix}
%a & c & 0 & \\
%b & a & c & & \cdots \\
%0 & b & a & \\
%& \vdots & & \ddots & c \\
%& & & b & a
%\end{pmatrix},
%\end{equation}
whose eigenvalues are known (see \cite{KULKARNI199963}):
\begin{equation}
    a - 2 \sqrt{bc} \cos{(k \pi / (n+1))}~, 
\end{equation}
where $k = 1, 2, \ldots, n$. In our case, $a = -2\sin{(-\frac{2 \pi j}{(\ell + 1)} + \varphi)}$, $b = c = \sin{(-\frac{2 \pi j}{(\ell + 1)} + \varphi)}$, so by directly using this result we see the eigenvalues of the Hessian (labeled by $k$) are given by
\begin{equation} \label{eigenvalues}
    2\sin{\Big(\frac{2 \pi j}{(\ell + 1)} - \varphi \Big)} \Big(1 - \cos{\big(k \pi /(\ell + 1) \big)}\Big)~, \hspace{1cm} k = 1, 2, \ldots, \ell~.
\end{equation}
Crucially, the signs of all eigenvalues for a given $\mathbf{X}^{(j)}$ are determined solely by the first factor in \eqref{eigenvalues} which allows us to read-off if the Hessian is positive (negative) definite for a given $\mathbf{X}^{(j)}$, and so conclude whether the corresponding $\mathbf{X}^{(j)}$ is a strict local minimum (maximum).

Finally, let us show that the latter minima (maxima) are actually global minima (maxima) of the problem. To see this consider equations \eqref{crit1}-\eqref{crit3}. We have shown the vanishing of all the sine factors on the right imply the  $Y^{(j)}_{i} = 2 \pi (A^{-1})_{ij}$ solutions. What happens when at least one of the sine factors on the left vanishes? Say e.g.\ $\sin{\Big(\frac{Y_{j+1}-Y_{j-1}+2\varphi}{2}\Big)} = 0$, which implies $Y_{j+1} - Y_{j-1} +2\varphi= 2 \pi n$ for some integer $n$. Then, take the $j$ and $j-1$ summands in \eqref{deformation-appendix} and notice that
\begin{align}
    \sin{(Y_{j + 1} - Y_{j} + \varphi)} + \sin{(Y_{j} - Y_{j - 1} + \varphi)} = \sin{(Y_{j - 1} - Y_{j} - \varphi)} + \sin{(Y_{j} - Y_{j - 1} + \varphi)} = 0~.
\end{align}
Thus, this implies that the deformation has at most $\ell - 1$ possibly non-vanishing terms in \eqref{deformation-appendix} instead of $\ell + 1$, and all bounded above by 1. Furthermore, at generic $\varphi$ it is straightforward to see that the greatest $\mathbf{X}^{(j)}$ maxima is bounded from below by $(\ell+1)\sin{(\frac{\pi}{2}-\frac{\pi}{(\ell + 1)})}$, and it is not difficult to see that
\begin{equation}
    (\ell+1)\sin{\bigg( \frac{\pi}{2}-\frac{\pi}{(\ell + 1)} \bigg)} > (\ell-1)~,
\end{equation}
for $\ell \geq 2$. So, the set of the $\mathbf{X}^{(j)}$ indeed contains the global maximum, and similar conclusions extend to the global minimum.

\section{Comments on Flows for Other Simply-Laced Algebras} \label{Other-Simply-Laced-Algebras}

In this appendix we comment on the relevant flows for the $D_{\ell}$, $E_{6}$ and $E_{7}$ algebras at level 1 where a free boson construction also exists. 

We start by quickly addressing the case of $D_{\ell}$ for $\ell \geq 8$ where only the deformation associated to the $\mathbf{w}_{1}$ (fundamental representation) is relevant. Since the corresponding representation is self-conjugate, the situation is analogous to the $SU(2)_{1}$ deformations of section \ref{su2case-subsection}, where there is only the cosine deformation
\begin{align} \label{Dlw1deformation}
    C^{D_{\ell}}_{\mathbf{w}_{1}} \sim  \cos{\big( \mathbf{w}_{1} \cdot \mathbf{X} \big)} &+ \sum_{j=2}^{\ell-2} \cos{\big( (\mathbf{w}_{j} - \mathbf{w}_{j-1}) \cdot \mathbf{X} \big)} \nonumber \\ &+ \cos{\big( (\mathbf{w}_{\ell-1} + \mathbf{w}_{\ell} - \mathbf{w}_{\ell-2}) \cdot \mathbf{X} \big)} + \cos{\big( (\mathbf{w}_{\ell} - \mathbf{w}_{\ell-1}) \cdot \mathbf{X} \big)} ~.
\end{align}
It is not difficult to see from the $D_{\ell}$ inverse Cartan matrix
\begin{equation} \label{DlinverseCartan}
(A^{-1})^{D_{\ell}}_{i,j} = \mathbf{w}_{i} \cdot \mathbf{w}_{j} = \frac{1}{2}
\begin{pmatrix}
2 & 2 & 2 & \ \ldots & 2 & 1 & 1 \\
2 & 4 & 4 & \ \ldots & 4 & 2 & 2 \\
2 & 4 & 6 & \ \ldots & 6 & 3 & 3 \\
. & . & . & \ \ldots & . & . & . \\
2 & 4 & 6 & \ \ldots & 2(\ell-2) & \ell-2  & \ell-2 \\
1 & 2 & 3 & \ \ldots & \ell-2 & \ell/2 & (\ell-2)/2 \\
1 & 2 & 3 & \ \ldots & \ell-2 & (\ell-2)/2 & \ell/2
\end{pmatrix}~,
\end{equation}
that this potential has doubly degenerate global maxima and minima.  Indeed, $\mathbf{X}^{(0)} = 0$ and $\mathbf{X}^{(1)} = 2 \pi \mathbf{w}_{1}$ are not related by translation over the root lattice and both attain the obvious global maximum of the deformation $C^{D_{\ell}}_{\mathbf{w}_{1}}\big|_{\mathrm{max}} = 2\ell$. Similarly, $\mathbf{X}^{(\ell-1)} = 2 \pi \mathbf{w}_{\ell-1}$ and $\mathbf{X}^{(\ell)} = 2 \pi \mathbf{w}_{\ell}$ both attain the obvious global minimum of the deformation.  Thus for $\ell \geq 8,$ the possible relevant flows result in degenerate ground states, i.e.\ a TQFT at long distances.  

We are thus reduced to analyzing a finite number of algebras. For $D_{\ell}$, $4 \leq \ell \leq 7$ the deformations associated to representations given by the weights $\mathbf{w}_{\ell-1}$ and $\mathbf{w}_{\ell}$ (corresponding to the chiral spinor representations) are now relevant ($\Delta_{\ell-1} = \Delta_{\ell} = \ell/4$) and can be considered in the RG flow. This will lift the degeneracies we have found for $\ell \geq 8$ and will allows us to realize SPT transitions.

For even $\ell$ the representations given by $\mathbf{w}_{\ell-1}$ and $\mathbf{w}_{\ell}$ are self-conjugate meaning that similarly as with \eqref{Dlw1deformation} we only have the corresponding cosine deformations $C^{D_{\ell}}_{\mathbf{w}_{\ell-1}}$, $C^{D_{\ell}}_{\mathbf{w}_{\ell}}$ (i.e. the sine deformations $S^{D_{\ell}}_{\mathbf{w}_{\ell-1}} = S^{D_{\ell}}_{\mathbf{w}_{\ell}} = 0$). For odd $\ell$ instead the representations given by $\mathbf{w}_{\ell-1}$ and $\mathbf{w}_{\ell}$ are conjugate to each other, implying $C^{D_{\ell}}_{\mathbf{w}_{\ell-1}} = C^{D_{\ell}}_{\mathbf{w}_{\ell}}$ and $S^{D_{\ell}}_{\mathbf{w}_{\ell-1}} = -S^{D_{\ell}}_{\mathbf{w}_{\ell}}$. So, the general deformations are:
\begin{equation} \label{Vleven}
    V^{D_{\ell}}_{\ell}(\alpha, \beta, \gamma) = \frac{\alpha}{2\ell} C^{D_{\ell}}_{\mathbf{w}_{1}} + \frac{\beta}{2^{\ell-1}} C^{D_{\ell}}_{\mathbf{w}_{\ell-1}} + \frac{\gamma}{2^{\ell-1}} C^{D_{\ell}}_{\mathbf{w}_{\ell}}~, \quad \mathrm{for \ \ell \ even}~,
\end{equation}
and
\begin{equation} \label{Vlodd}
    V^{D_{\ell}}_{\ell}(\alpha, \beta, \gamma) = \frac{\alpha}{2\ell} C^{D_{\ell}}_{\mathbf{w}_{1}} + \frac{\beta}{2^{\ell-1}} C^{D_{\ell}}_{\mathbf{w}_{\ell-1}} + \frac{\gamma}{2^{\ell-1}} S^{D_{\ell}}_{\mathbf{w}_{\ell-1}}~, \quad \mathrm{for \ \ell \ odd}~.
\end{equation}
The denominator for each deformation has been chosen as the dimension of the corresponding representation for convenience.

\begin{table}[t]
\centering
 \begin{tabular}{|lllll|}
 \hline
  & $V^{D_{4}}_{4}(\alpha, \beta, \gamma)$ & $V^{D_{5}}_{5}(\alpha, \beta, \gamma)$ & $V^{D_{6}}_{6}(\alpha, \beta, \gamma)$ & $V^{D_{7}}_{7}(\alpha, \beta, \gamma)$ \\ [0.5ex] 
 \hline\hline
 $\mathbf{X} = 0$ & $(\alpha + \beta + \gamma)$ & $(\alpha + \beta)$ & $(\alpha + \beta + \gamma)$ & $(\alpha + \beta)$ \\ 
 $\mathbf{X} = 2 \pi \mathbf{w}_{1}$ & $(\alpha - \beta - \gamma)$ & $(\alpha - \beta)$ & $(\alpha - \beta - \gamma)$ & $(\alpha - \beta)$ \\
  $\mathbf{X} = 2 \pi \mathbf{w}_{\ell-1}$ & $(-\alpha + \beta - \gamma)$ & $(-\alpha + \gamma)$ & $(-\alpha - \beta + \gamma)$ & $(-\alpha - \gamma)$ \\
    $\mathbf{X} = 2 \pi \mathbf{w}_{\ell}$ & $(-\alpha - \beta + \gamma)$ & $(-\alpha - \gamma)$ & $(-\alpha + \beta - \gamma)$ & $(-\alpha + \gamma)$ \\
 \hline
 \end{tabular}
 \caption{Table of Extrema  of $V^{D_{\ell}}_{\ell}(\alpha, \beta, \gamma)$ at $X^{(j)} = 2 \pi \mathbf{w}_{j}$ for $4 \leq \ell \leq 7$.}
\label{table:3}
\end{table}

In order to see that for at least some range of parameters the deformations \eqref{Vleven} and \eqref{Vlodd} lead to SPT transitions we can take the following set-up: Take $\alpha > 0$ and $\beta = \gamma = 0$. By the same analysis as in the $\ell \geq 8$ case we know there are two degenerate global maxima at $(\mathbf{X}^{(0)},\mathbf{X}^{(1)})$ and two degenerate global minima at $(\mathbf{X}^{(\ell-1)},\mathbf{X}^{(\ell)})$. These maxima and minima saturate the bound $|C^{D_{\ell}}_{\mathbf{w}_{1}}| \leq 2 \ell$. Using this fact and the condition for $C^{D_{\ell}}_{\mathbf{w}_{1}}$ to be extremized it is possible to show that no other critical points exist which saturate such bound. Furthermore, from the condition for $V^{D_{\ell}}_{\ell}(\alpha, \beta, \gamma)$ to be extremized it is also possible to see that the $(\mathbf{X}^{(0)},\mathbf{X}^{(1)})$ and $(\mathbf{X}^{(\ell-1)},\mathbf{X}^{(\ell)})$ critical points exist for any values of $(\alpha, \beta, \gamma)$. Thus, if we now perturb the $\alpha > 0, \, \beta = \gamma = 0$ set-up with some small values of $\beta$ and $\gamma$ the degeneracies will be lifted and we will find trivially gapped phases, at least for some region around $\beta = \gamma = 0$. The values of these extrema for $4 \leq \ell \leq 7$ are shown in Table \ref{table:3}, from where we can check that the degeneracies are indeed lifted. From Table \ref{table:3} it is also straightforward to see that any SPT transition can be realized depending on the ratios of $\alpha$, $\beta$, $\gamma$, and comparing partition functions of different flows with background fields for the respective Cartan subalgebras. That is:
\begin{equation} \label{quotient-of-partition}
    \lim_{|\alpha_{1,2}|, |\beta_{1,2}|, |\gamma_{1,2}| \rightarrow \infty} \frac{Z[A, \alpha_{1}, \beta_{1}, \gamma_{1}]}{Z[A, \alpha_{2}, \beta_{2}, \gamma_{2}]} = \exp{\bigg( 2 \pi i (A^{-1})^{D_{\ell}}_{i,j} \Big(\frac{1}{2\pi}\int dA^{j}\Big) \bigg)}~,
\end{equation}
for any $i=1,\ell-1,\ell$, at least for some region in $\beta$ and $\gamma$ as long as their ratio respect to $\alpha$ is small enough. We have written this quotient of partition functions in terms of the $D_{\ell}$ inverse Cartan matrix \eqref{DlinverseCartan} from where it is direct to verify that for $i=0,1,\ell-1,\ell$ (where we do $(A^{-1})_{0,j} = 0$) the expected group cohomology $\mathbb{Z}_{2} \times \mathbb{Z}_{2}$ for $\ell$ even, or $\mathbb{Z}_{4}$ for $\ell$ odd is reproduced. That is, just as for the $A_{\ell}$ algebras, the inverse Cartan matrix already contains enough information to realize the group cohomology classification of the SPT phases, and the free boson construction provides a concrete physical realization leading to the appropriate vacua.

A similar analysis can be performed for the exceptional $E_{6}$ and $E_{7}$ algebras. In the conventions of \cite{DiFrancesco:639405}, $E_{7}$ has a single relevant deformation $C_{\mathbf{w}_{6}}^{E_{7}}$ corresponding to the $56$-dimensional, self-conjugate representation given by $\mathbf{w}_{6}$. Again we see that the global maximum (minimum) $C_{\mathbf{w}_{6}}^{E_{7}} = + 56$ ($-56$) is attained at $\mathbf{X}^{(0)} = 0$ ($\mathbf{X}^{(6)} = 2 \pi \mathbf{w}_{6}$). Following the same steps as above for $D_{\ell}$, we can check no other critical points exist with these global maxima and minima and thus the $\mathbb{Z}_{2}$ SPT transition can be realized.

\begin{table}[t]
\centering
 \begin{tabular}{|ll|}
 \hline
  & \hspace{0.5cm} $V^{E_{6}}(\lambda,\theta)$ \\ [0.5ex] 
 \hline\hline
 $\mathbf{X} = 0$ & \hspace{0.6cm} $\lambda \cos{\theta}$ \\ 
 $\mathbf{X} = 2 \pi \mathbf{w}_{1}$ & \, $\lambda \cos{(\theta + \frac{2\pi}{3})}$ \\
 $\mathbf{X} = 2 \pi \mathbf{w}_{5}$ & \, $\lambda \cos{(\theta - \frac{2\pi}{3})}$ \\
 \hline
 \end{tabular}
 \caption{Table of Extrema of $V^{E_{6}}(\lambda, \theta)$ at $X^{(j)} = 2 \pi \mathbf{w}_{j}$.}
\label{table:4}
\end{table}

Finally, for $E_{6}$ there are two representations given by $\mathbf{w}_{1}$ and $\mathbf{w}_{5}$ (again in the conventions of \cite{DiFrancesco:639405}), which are conjugate to each other and leading to relevant deformations. So, we set up the deformation
\begin{equation}
    V^{E_{6}}(\lambda, \theta) = \lambda \Big(\frac{\cos{\theta}}{27} C^{E_{6}}_{\mathbf{w}_{1}} - \frac{\sin{\theta}}{27} S^{E_{6}}_{\mathbf{w}_{1}} \Big),
\end{equation}
which satisfies $|V^{E_{6}}(\lambda, \theta)| \leq \lambda$. As before, there are extrema $\mathbf{X}^{i} = 2\pi \mathbf{w}_{i}$ for the fundamental weights associated to integrable representations at level 1 (i.e. $i=0,1,5$). The values of the deformation at these critical points are shown in Table \ref{table:4}. At $\theta = 0$ and $\theta = \pm 2\pi/3$ we see the bound is saturated by one of these critical points, and we can use the same steps as in the $D_{\ell}$ case above to show that no other points can be found that saturate the bound. Thus, performing flows for some finite regions around $\theta = 0, \pm 2\pi/3$, a quotient analogous to \eqref{quotient-of-partition} can be found for any $i=0,1,5$ with the inverse Cartan matrix replaced by that of $E_{6}$. The corresponding rows of the inverse Cartan matrix realize the corresponding $\mathbb{Z}_{3}$ group cohomology just as in the previous cases.

%In \cite{2017}, the bound $c > \log_{2}(d)$ was proposed. Let us recall the rationale behind this proposal

%The modular S-matrix in the $G(2)_{1}$ case is given by
%\begin{equation}
%    S = \sqrt{\frac{4}{5}} \begin{pmatrix}
%\sin{(\frac{\pi}{5})} & \sin{(\frac{3\pi}{5})}  \\
%\sin{(\frac{3\pi}{5})} & -\sin{(\frac{\pi}{5})} 
%\end{pmatrix}
%\end{equation}

%Effectively then we are dealing with the $G(2)_{1}$ WZW CFT with boundary conditions described by the boundary states $|0\rrangle$ and $|2/5\rrangle$, and from \eqref{Cardy-OpenPartition} we know only the $\chi_{2/5}$ character will contribute. But the corresponding representation has dimension $\mathrm{dim}(\mathbf{w}_{2}) = 7$ (using the notation of \cite{DiFrancesco:639405}), and the central charge of the $G(2)_{1}$ WZW is $c=14/5$. Plugging in directly, we find:
%\begin{equation}
%    c - \log_{2}(d) = \frac{14}{5} - \log_{2}(7) = -0.00735\ldots < 0,
%\end{equation}
%which directly contradicts the rationale used in \cite{2017} to motivate the $c > \log_{2}(d)$ bound for SPT transitions.

\section{A Bound on $c$ for Irrational Theories?} \label{appendixirrationaltheories}

Throughout this work we have mainly considered rational theories where boundary conditions are fairly well-understood and systematic analysis is possible.  By contrast, in irrational CFTs there is no general method known to classify boundary conditions and this ignorance presents a challenge towards generalizing our results (in particular \eqref{first-bound}) to this broader context.

Recall that to obtain \eqref{first-bound} we used in \eqref{inequalityfromCardyCond} that the modular $S$ matrix is unitary, and rationality ensured the finiteness of $\sum_{j} |n^{j}_{AB}|$. In an irrational theory the formal sum $\sum_{j} |n^{j}_{AB}|$ is in general infinite.  However, cancellation of phases of the modular $S$ matrix may still make $|\sum_{j} n^{j}_{AB}S^{-1}_{ji}|$ finite. However, the details of such cancellation, if it happens, are in principle theory dependent.

As a first step towards studying the problem of irrational theories, in this appendix we focus on the particular example of a free boson compactified on a circle of arbitrary radius $R$, where conformal boundary conditions have been classified in \cite{Gaberdiel:2001xm, Gaberdiel:2001zq, Janik:2001hb, Friedan-Unpublished}.

Following equation \eqref{Introducing-Resolutions-of-Identity} notice that a sufficient condition to ensure a bound is that, when introducing the resolution of the identity in terms of the Virasoro states $ \sum_{\mathcal{O}} |\mathcal{O} \rangle \langle \mathcal{O} |$, the inner products are bounded
\begin{equation} \label{O-independent}
    | \langle \mathcal{O} | A \} | \leq \mathcal{N}_{A} \,~,
\end{equation}
for some $\mathcal{N}_{A}$ independent of $\mathcal{O}$.

We will check directly that this is the case for the boundary states considered \cite{Gaberdiel:2001xm, Gaberdiel:2001zq, Janik:2001hb, Friedan-Unpublished}. To set up notation recall the partition function for the free boson compactified in a circle of radius $R$ is given by:
\begin{equation} \label{FreeBosonPartitionFunction}
    Z(\tau, \overline{\tau}) = \frac{1}{|\eta(\tau)|^{2}} \sum_{k,w \in \mathbb{Z}} q^{\frac{1}{2}\big(\frac{k}{R} + \frac{w R}{2}\big)^{2}} \overline{q}^{\frac{1}{2}\big(\frac{k}{R} - \frac{w R}{2}\big)^{2}}~,
\end{equation}
and let us denote correspondingly the (normalized) highest-weight states with momentum $k$ and winding $w$ by $| (k,w) \rangle$.

When $R$ is an irrational multiple of the self-dual radius $R_{\mathrm{self-dual}} = \sqrt{2}$ there are two sets of boundary states (see \cite{Janik:2001hb}). The first set corresponds to the usual Dirichlet/Neumann states:
\begin{align}
    | D(x_{0}) \} &= \sqrt{\frac{1}{R}} \Bigg[ \sum_{J=0}^{\infty} | J \rrangle + \sum_{k \in \mathbb{Z}/\{0\}} e^{i k x_{0} / R}  | (k,0)\rrangle \Bigg]~, \label{Dirichlet} \\[0.25cm]
    | N(\tilde{x}_{0}) \} &= \sqrt{\frac{R}{2}} \Bigg[ \sum_{J=0}^{\infty} (-1)^{J} | J \rrangle + \sum_{w \in \mathbb{Z}/\{0\}} e^{ i R w \tilde{x}_{0} / 2}  | (0,w)\rrangle \Bigg]~. \label{Neumann}
\end{align}
Here $|(k,0)\rrangle$ \big($|(0,w)\rrangle$\big) denotes the Ishibashi state associated to the $(k,0)$ \big($(0,w)$\big) primary. The $|J\rrangle$ labels the Virasoro Ishibashi states for the Virasoro irreducible representations appearing in the decomposition $\chi^{U(1)}_{k=w=0}(q) = \sum_{J=0}^{\infty} \chi^{Vir}_{J^{2}}(q)$ of the trivial primary $k=w=0$.

Apart from these states there is an additional, second set of states, first suggested by Friedan \cite{Friedan-Unpublished}. As shown in \cite{Janik:2001hb}, these are given by
\begin{equation} \label{Friedan-states}
    |x\} = \sum_{J} P_{J}(x) | J \rrangle~,
\end{equation}
up to an overall normalization, where $-1<x<1$, and $P_{J}(x)$ is a Legendre polynomial. 

Now, since a Virasoro state has a non-zero inner product with at most one Ishibashi state, when we take the inner product of a Virasoro state and any of the previous boundary states we will obtain either a phase (times an overall normalization), or a Legendre polynomial coefficient which satisfies $|P_{J}(x)| \leq 1$ for  $-1<x<1$. Therefore, we see \eqref{O-independent} indeed holds with a right-hand side that is independent of the Virasoro state in question.

If instead $R$ is the self-dual radius $R = R_{\mathrm{self-dual}} = \sqrt{2}$ the first set of boundary states \eqref{Dirichlet}-\eqref{Neumann} remains, but now the boundary state \eqref{Friedan-states} is no longer defined.  Instead there is another set of boundary states parameterized by a $SU(2)$ element $g$. The explicit form of this set of states is given by
\begin{equation} \label{Gaberdiel-like-boundarystate}
    | g \} = \frac{1}{2^{1/4}} \sum_{j,m,n} D^{j}_{m,n}(g)|j,m,n\rrangle~,
\end{equation}
where $|j,m,n\rrangle$ for $j$ non-negative half-integer and $m,n$ half-integers such that $|m|,|n| \leq j$, label Virasoro Ishibashi states of the theory (see \cite{Gaberdiel:2001xm, Gaberdiel:2001zq} for more details). For completeness, the explicit expression for $D^{j}_{m,n}(g)$ is
\begin{align}
    & D^{j}_{m,n}(g) = \sum_{l = \mathrm{max}(0,n-m)}^{\mathrm{min}(j-m,j+n)} \frac{[(j+m)!(j-m)!(j+n)!(j-n)!]^{1/2}}{(j-m-l)!(j+n-l)!l!(m-n+l)!} \nonumber \\ & \hspace{7cm} \times (a)^{j+n-l}(a^{*})^{j-m-l}(b)^{m-n+l}(-b^{*})^{l}~,
\end{align}
where we have taken
\begin{equation}
    g = \begin{pmatrix}
a & b \\
-b^{*} & a^{*} 
\end{pmatrix}~.
\end{equation}
This last set of states is, actually, the set of all possible Cardy states including all possible $\Omega$-twistings \cite{Gaberdiel:2001xm} (see \eqref{extendedgluing} for definition of $\Omega$). Importantly, the matrix $D^{j}(g)$ with components written above is unitary, from which we immediately know that the entries are bounded in absolute value by $1$. Thus, arguing similarly as above we also have \eqref{O-independent} with a right hand side independent of $\mathcal{O}$ for any $g$. 

Finally, when $R$ is some rational multiple of the self-dual radius $R = \frac{M}{N} R_{\mathrm{self-dual}}$ with $M,N$ integers, we can perform an analogous analysis to the case $R = R_{\mathrm{self-dual}}$. The main point to stress is that although the explicit expression analogous to \eqref{Gaberdiel-like-boundarystate} is different (see Eqn. (4.7) in \cite{Gaberdiel:2001zq}), the coefficients are still universally bounded by the fact that they are also given in terms of entries of unitary matrices, so similar conclusions as the case $M=N=1$ follow.

\bibliographystyle{JHEPmod}
\bibliography{references}

\end{document}